\documentclass[12pt]{iopart}
\usepackage{iopams}  

\usepackage[english]{babel} 
\usepackage[utf8]{inputenc} 
\usepackage{textcomp} 
\usepackage{booktabs} 
\usepackage{multirow}
\usepackage{array}
\usepackage{subfig} 
\usepackage{caption}
\usepackage{hyperref}
\usepackage{graphicx}
\usepackage{xcolor}
\usepackage{upgreek}
\usepackage{ulem}

\begin{document}

\title{Exploration of co-sputtered Ta$_2$O$_5$-ZrO$_2$ thin films for gravitational-wave detectors}

\author{M Abernathy$^{1,2}$, A Amato$^{3}$, A Ananyeva$^{4}$, S Angelova$^{5}$, B Baloukas$^{6}$, R Bassiri$^{7}$, G Billingsley$^{4}$, R Birney$^{5,8}$, G Cagnoli$^{3}$, M Canepa$^{9,10}$, M Coulon$^{11}$, J Degallaix$^{11}$, A Di Michele$^{12}$, M A Fazio$^{13}$, M M Fejer$^{7}$, D Forest$^{11}$, C Gier$^{5}$, M Granata$^{11}$, A M Gretarsson$^{14}$, E M Gretarsson$^{15}$, E Gustafson$^{4}$, E J Hough$^{16}$, M Irving$^{17}$, \'{E} Lalande$^{17}$, C L\'{e}vesque$^{17}$, A W Lussier$^{17}$, A Markosyan$^{7}$, I W Martin$^{16}$, L Martinu$^{6}$, B Maynard$^{18}$, C S Menoni$^{13,19}$, C Michel$^{11}$, P G Murray$^{16}$, C Osthelder$^{4}$, S Penn$^{18}$, L Pinard$^{11}$, K Prasai$^{7}$, S Reid$^{5}$, R Robie$^{16}$, S Rowan$^{16}$, B Sassolas$^{11}$, F Schiettekatte$^{17}$, R Shink$^{17}$, S Tait$^{16}$, J Teillon$^{11}$, G Vajente$^{4}$, M Ward$^{17}$, L Yang$^{19}$}
\address{$^{1}$Department of Physics, American University, Washington, DC, USA}
\address{$^{2}$Johns Hopkins University Applied Physics Laboratory, Laurel, MD, USA}
\address{$^{3}$Universit\'{e} de Lyon, Universit\'{e} Claude Bernard Lyon 1, CNRS, Institut Lumi\`{e}re Mati\`{e}re, F-69622 Villeurbanne, France}
\address{$^{4}$LIGO Laboratory, California Institute of Technology, Pasadena, CA, USA}
\address{$^{5}$SUPA, Department of Biomedical Engineering, University of Strathclyde, Glasgow, UK}
\address{$^{6}$\'{E}cole Polytechnique de Montr\'{e}al, Montr\'{e}al, Qu\'{e}bec, Canada}
\address{$^{7}$Edward L. Ginzton Laboratory, Stanford University, Stanford, CA 94305, USA}
\address{$^{8}$SUPA, Institute of Thin Films, Sensors and Imaging, University of the West of Scotland, Paisley, UK}
\address{$^{9}$OPTMATLAB, Dipartimento di Fisica, Universit\`{a} di Genova, Via Dodecaneso 33, 16146 Genova, Italy}
\address{$^{10}$INFN, Sezione di Genova, Via Dodecaneso 33, 16146 Genova, Italy}
\address{$^{11}$Laboratoire des Mat\'{e}riaux Avanc\'{e}s, Institut de Physique des 2 Infinis de Lyon, CNRS/IN2P3, Universit\'{e} de Lyon, Universit\'{e} Claude Bernard Lyon 1, F-69622 Villeurbanne, France}
\address{$^{12}$Dipartimento di Fisica e Geologia, Universit\`{a} degli Studi di Perugia, Via Pascoli, 06123 Perugia, Italy}
\address{$^{13}$Department of Electrical and Computer Engineering, Colorado State University, Fort Collins, CO, USA}
\address{$^{14}$Embry-Riddle Aeronautical University, 3700 Willow Creek Rd., Prescott, AZ 86301, USA}
\address{$^{15}$Wyant College of Optical Sciences, University of Arizona, 1630 E. University Blvd., Tucson, AZ 85721, USA}
\address{$^{16}$SUPA, Institute for Gravitational Research, University of Glasgow, Glasgow, UK}
\address{$^{17}$Universit\'{e} de Montr\'{e}al, Montr\'{e}al, Qu\'{e}bec, Canada}
\address{$^{18}$Department of Physics, Hobart and William Smith Colleges, Geneva, NY, USA}
\address{$^{19}$Department of Chemistry, Colorado State University, Fort Collins, CO, USA}

\ead{m.granata@in2p3.fr, vajente@caltech.edu}

\begin{abstract}
We report on the development and extensive characterization of co-sputtered tantala-zirconia thin films, with the goal to decrease coating Brownian noise in present and future gravitational-wave detectors. We tested a variety of sputtering processes of different energies and deposition rates, and we considered the effect of different values of cation ratio $\eta =$ Zr/(Zr+Ta) and of post-deposition heat treatment temperature $T_a$ on the optical and mechanical properties of the films. Co-sputtered zirconia proved to be an efficient way to frustrate crystallization in tantala thin films, allowing for a substantial increase of the maximum annealing temperature and hence for a decrease of coating mechanical loss. The lowest average coating loss was observed for an ion-beam sputtered sample with $\eta = 0.485 \pm 0.004$ annealed at 800 $^{\circ}$C, yielding $\overline{\varphi} = 1.8 \times 10^{-4}$. All coating samples showed cracks after annealing. Although in principle our measurements are sensitive to such defects, we found no evidence that our results were affected. The issue could be solved, at least for ion-beam sputtered coatings, by decreasing heating and cooling rates down to 7 $^{\circ}$C/h. While we observed as little optical absorption as in the coatings of current gravitational-wave interferometers (0.5 parts per million), further development will be needed to decrease light scattering and avoid the formation of defects upon annealing. 
\end{abstract}

\vspace{2pc}
\noindent{\it Keywords}: xxx, yyy, zzz

%
\section{Introduction}
The detection of gravitational-wave (GW) signals from astrophysical systems \cite{gw150914} has ushered in a new era in astronomy \cite{multimessenger, gw170817}, allowing the study of systems previously invisible. The current generation of GW interferometric detectors are sensitive enough to allow a confident candidate event about once per week \cite{gwtc-1, gwtc-2}. Increasing the sensitivity of the detectors is important not only to increase the detection rate but also to improve the signal to noise ratio and sky localization of events, allowing for a more precise probing of the properties of nuclear matter in neutron stars \cite{ns} and of the fundamental physics of gravitation and general relativity \cite{test-gr}. 

Ironically, the sensitivity of current detectors to signals from very large and massive astrophysical systems is limited by random noise in microscopic systems, that is, fluctuations resulting from the internal mechanical dissipation in the material constituting the mirrors used as test masses \cite{thermal-noise}. In the most sensitive region of the detection band of Advanced LIGO \cite{aLIGO} and Advanced Virgo \cite{aVirgo}, between 50 and 300 Hz, the sensitivity is limited by an equal contribution of quantum noise \cite{quantum-noise} and thermal noise, the latter being dominated by coating Brownian noise \cite{Fejer:16}. Indeed, all GW detectors currently operating are laser interferometers with km-scale optical resonators \cite{current-gw}, where the test masses used to probe the space time metric fluctuations are high-quality fused silica mirrors with high-reflection (HR) dielectric coatings on the front surfaces \cite{Degallaix19,Pinard:17,coatings}. Of all the sources of thermal noise in the mirrors, the Brownian motion of particles in the coating layers is the limiting factor and hence needs to be reduced for future sensitivity upgrades \cite{aplus, einstein-telescope, cosmic-explorer}.

The dielectric coatings used in the current generation of GW detectors are Bragg reflectors \cite{multilayer-coatings} made of alternate ion-beam sputtered (IBS) thin layers of silica (SiO$_2$) and tantala-titania (Ta$_2$O$_5$-TiO$_2$). Silica is the low refractive index material ($n = 1.45$ at 1064 nm), and the tantala-titania is the high refractive index material ($n = 2.07$ at 1064 nm) \cite{Granata20}. These materials and their synthesis by various methods have been extensively investigated in the context of optical materials in general and interferometric filters in particular \cite{multilayer-coatings}, including for example their mechanical properties \cite{Klemberg-Sapieha2004, Cetinorgu2009,Granata20}, showing that their performance is intimately related to parameters such as their internal stress and growth parameters \cite{Granata20}.

The detection of GW signals is performed by monitoring the change in the relative distance between the mirrors, measured as a change in the phase of the laser beams \cite{fundamentals-gw-detection}. Therefore, any fluctuation in the thickness or in the optical properties of the coating materials acts like a spurious displacement noise. Thermo-refractive and thermo-elastic fluctuations are not relevant \cite{thermo-optic}, and the dominant source of phase noise is due to the Brownian motion of all the layers. The fluctuation dissipation theorem \cite{FDT} links Brownian noise to the rate of energy dissipation in the coating material due to internal friction. The amount of friction is measured by the material loss angle $\varphi$, that is the fraction of energy lost per cycle when the material is driven by a sinusoidal external force. In the coating materials used so far, the worst offender is tantala-titania, with measured values $(2.3-3.4) \times 10^{-4}$ \cite{gras18,Granata20}. The silica layers have one order of magnitude lower loss angle, $(2.3-4.0) \times 10^{-5}$ \cite{Penn_2003,Granata20}. 

The goal for the GW detector upgrades in the next 2-5 years is to roughly double the sensitivity to GW inspiral signals \cite{aplus}, which translates to an improvement of the same factor of the sensitivity in the $100$ Hz region. In Advanced LIGO, for instance, this sets a target of $6.6 \times 10^{-21} $ m$/\sqrt{\mbox{Hz}}$ at 100 Hz in terms of coating Brownian noise, to be compared with the current estimated level of $1.3 \times 10^{-20}$ m$/\sqrt{\mbox{Hz}}$ \cite{o3-detector-paper}. The Brownian noise amplitude spectral density scales with the square root of the total loss angle of the HR coating \cite{Harry_2002}. Theerfore the upgrade goal translates into improving the high refractive index material loss angle by at least of a factor 4.  Marginal or no improvement is necessary for the low refractive index material. For this reason, most of the coating research in the Virgo and LIGO collaborations focused on finding an alternative material to replace the tantala-titania layers \cite{coatings,coatingR&Dreview}, while keeping the same outstanding optical properties (absorption, scattering, etc.) of the HR coatings. The main line of research explores the co-sputtering of other oxides with tantala, following the success of co-sputtered tantala-titania coatings \cite{Harry2006} which allowed a reduction of the coating loss angle by about 25\% \cite{Granata20} with respect to tantala.

Beginning in 2012, one of the authors (S. P.) characterized a series of zirconia-based coatings. That investigation was motivated by zirconia (ZrO$_2$) being a high-index material ($n \ge 2$ at 1064 nm, like tantala) possessing a similar molecular structure and physical parameters to silica. However, unlike silica, when treated at high temperatures, zirconia will crystallize unless stabilized. Tewg et al. \cite{Tewg2005} showed that the addition of zirconia to tantala thin films allowed for a higher maximum heating temperature, before the onset of crystallization. Post-deposition heat treatment, referred to as {\it annealing} hereafter, is a standard procedure to decrease the internal stress, the optical absorption and the mechanical loss of coatings \cite{Granata20}. As a general empirical rule, the higher the annealing peak temperature, the lower the resulting coating mechanical loss. In the work of Tewg et al., the annealing had a duration of 10 minutes, while typical annealing times for HR coatings used in GW detectors are of the order of 10 hours \cite{Granata20}. Indeed, higher annealing temperatures (up to 900 $^\circ$C) and duration would be beneficial also for the silica layers, decreasing their mechanical loss \cite{Amato18,Granata18} as a result of structural relaxation, leading to HR coatings with overall lower loss angle and hence lower thermal noise.

This paper focuses on a collaborative work, carried out by different groups of the LIGO and Virgo Collaborations to produce and characterize co-sputtered tantala-zirconia thin films, for possible use as high refractive index layers for the new sets of mirrors of the Advanced LIGO+ and Advanced Virgo+ sensitivity upgrade projects. We considered the effect of adding various amounts of zirconia, and measured the optical and mechanical properties of the films as functions of their chemical composition and annealing temperature. As coating mechanical loss depends in general on the nature of the growth technique, we tested a variety of sputtering processes of different energies and deposition rates. A detailed description of the coating samples is given in Section \ref{SECT_samples}. Different techniques and methodologies have been used to characterize the optical and mechanical properties of the coating samples, described with their results in Sections \ref{SECT_opt} and \ref{SECT_mech}. In Sections \ref{SECT_mech} and \ref{SECT_annealing} we show that the main effect of co-sputtering zirconia with tantala is a significant increase of the maximum annealing temperature $T_a$, hence a reduction of coating mechanical loss. The main issues we encountered are cracking of all annealed single-layer samples and bubble-like defects in annealed IBS HR coatings. Later we found that cracking could be avoided by drastically slowing down the annealing heating and cooling rates of the samples. Bubble-like defects, however, remain a pending issue to date. In Section \ref{SECT_Raman} we present the results of Raman spectroscopy measurements, and discuss the possibility of using Raman peaks as a probe of the state of relaxation or a predictor of the loss angle in a material. Finally, we discuss the results and their implications in Section \ref{SECT_discussion}.
%
\section{Deposition methods and film composition}
\label{SECT_samples}
The co-sputtered tantala-zirconia coating samples used for this work were produced by one industrial manufacturer and several academic research groups, using different growth techniques:
\begin{itemize}
	\item ion-beam sputtering (IBS), at MLD Technologies (MLD) and Laboratoire des Mat\'eriaux Avanc\'es (LMA)
	\item electron-cyclotron-resonance ion-beam deposition (ECR-IBD), at University of Strathclyde (UoS)
	\item reactive biased-target deposition (RBTD), at Colorado State University (CSU), Fort Collins
	\item high-power impulse (HiPI) and radio-frequency (RF) magnetron sputtering (MS), at Universit\'e de Montr\'eal and Polytechnique Montréal (UMP).
\end{itemize}
Acronyms of growth techniques and groups are used throughout this paper to identify the samples. The measured atomic cation ratio $\eta$ = Zr/(Zr+Ta) and density $\rho$ of all samples considered in this work are summarized in Table \ref{TAB_samples} and plotted and discussed in Section \ref{SECT_opt_compar}. Details on the different growth processes, sample specifications and characterizations follow below.
\begin{table*}[!ht]
	\begin{center}
		\begin{tabular}{cccc}
			\hline
			Sample & Process & $\eta$ [at.\%] & $\rho$ [g/cm$^3$]\\
			\hline
            MLD2014 & IBS & $48.5 \pm 0.4$\\
            MLD2018 & IBS & $50.2 \pm 0.6$\\
            LMA & IBS & 19 $\pm$ 1 & 6.9 $\pm$ 0.1\\
            UoS & ECR-IBD & $34.0 \pm 0.5$\\
            UMP 551 & RF-MS & 41 $\pm$ 3 & 6.1 $\pm$ 0.2\\ 
            UMP 554 & RF-MS & 43 $\pm$ 3 & 5.9 $\pm$ 0.2\\ 
            UMP 658 & HiPIMS & 24 $\pm$ 3 & 7.0 $\pm$ 0.2\\ 
            UMP 659 & HiPIMS & 24 $\pm$ 2 & 7.1 $\pm$ 0.2\\ 
            UMP 542 & HiPIMS & 7 $\pm$ 2 & 7.3 $\pm$ 0.2\\ 
            UMP 664 & HiPIMS & 7 $\pm$ 2 & 7.1 $\pm$ 0.2\\ 
            UMP 678 & HiPIMS & 47 $\pm$ 2 & 6.6 $\pm$ 0.2\\ 
            UMP 680 & HiPIMS & 47 $\pm$ 2 & 6.5 $\pm$ 0.2\\ 
            CSU II & RBTD & 23 $\pm$ 1 & 7.1$^{(*)}$\\
            CSU III & RBTD & 54 $\pm$ 3 & 7.1$^{(*)}$\\
			\hline
		\end{tabular}
	\end{center}
	\caption{Measured atomic cation ratio $\eta$ = Zr/(Zr+Ta) and density $\rho$ of co-sputtered Ta$_2$O$_5$-ZrO$_2$ thin films. RF-MS samples had Ta deposited by RF-MS and Zr deposited by DC-MS, HiPIMS samples had Ta deposited by HiPIMS and Zr deposited by RF-MS. Density values are plotted and discussed in Section \ref{SECT_opt_compar}, where correlation between oxygen concentration, density and refractive index are considered. $^{(*)}$Estimated from component values and concentration.}
	\label{TAB_samples}
\end{table*}

\subsection{IBS}

Ion-beam-sputtered films where produced by MLD Technologies and by the LMA group. Details are described below.

\subsubsection{MLD Technologies --}
\label{SECT_MLDsamples}
The provider is a manufacturer of high performance and specialty optical coatings located in Mountain View, CA, with coating facilities in Eugene, OR. The company deposited the coatings using standard IBS, where a beam of Ar ions is directed at a metal target and the sputtered metal atoms are oxidized in a low density O$_2$ atmosphere before striking the substrate. For all deposition runs, MLD coated: (i) thin super-polished disk substrates of Corning 7980 fused silica that were 76.2 mm in diameter and 2.5-mm thick; (ii) a $\sim$100-$\mu$m thick fused-silica cantilever welded to a 10 mm $\times$ 10 mm $\times$ 3 mm clamping block; (iii) and a number of 35 mm $\times$ 5 mm $\times$ $\sim$65 $\mu$m silicon cantilevers with a 10 mm $\times$ 5 mm $\times$ $\sim$550 $\mu$m clamping block. As is standard practice, each run included fused-silica witness samples with 25.4 mm in diameter that were used to test the coating composition and thickness. 

MLD conducted 2 runs, whose samples are referenced as {\it MLD2014} and {\it MLD2018}, of identical specifications: the design goal was a single 500-nm thick tantala-zirconia coating with atomic cation ratio $\eta = \mathrm{Zr}/\left(\mathrm{Zr} + \mathrm{Ta}\right) \sim 36$\%, as specified by Tewg et al.~\cite{Tewg2005} for achieving amorphous coatings stable to peak annealing temperatures of 800 $^\circ$C, where the ratio was intended to denote atomic number density. MLD performed post-deposition energy-dispersive X-ray spectroscopy (EDX) measurements that reported cation ratios in agreement with that goal. However, while Tewg et al. reported atomic number densities, MLD reported ratios were later learned to be molecular number densities. Moreover, a reanalysis of the EDX data showed that the $\eta$ was $48.5\pm 0.4$\% for MLD2014 and $50.2 \pm 0.6$\% for MLD2018. Thus the actual molecular number densities were $\mathrm{ZrO_2} = 65.3 \pm 0.4\%$ and $\mathrm{Ta_2O_5} = 34.7 \pm 0.4\%$ for MLD2014 and $\mathrm{ZrO_2} = 66.8 \pm 0.6\%$ and $\mathrm{Ta_2O_5} = 33.2 \pm 0.6\%$ for MLD2018. MLD also reported coating thicknesses of 590 nm for MLD2014 and 483 nm for MLD2018.

After coating deposition, to test the effect of annealing on coating mechanical loss, the fused-silica disk and the silicon cantilevers were progressively annealed with steps of increasingly high temperatures: 400, 600 and 700 $^\circ$C for one set of samples; 300, 400, 600, 700, and 800$^\circ$C for a second set of samples.

\subsubsection{LMA --}
Co-sputtered Ta$_2$O$_5$-ZrO$_2$ coatings were produced with the custom-developed so-called {\it Grand Coater} (with a volume of 10 m$^3$) used to coat the mirrors of the Advanced LIGO, Advanced Virgo and KAGRA detectors \cite{Pinard:17,Degallaix19,Granata20}. Prior to deposition, the base pressure was less than 10$^{-7}$ mbar. Ar gas was fed into the sputtering source while O$_2$ gas was fed directly into the vacuum chamber, for a total working pressure of $2 \times 10^{-4}$ mbar. Energy and current of the sputtering Ar ions were of the order of 1 keV and 0.1 A, respectively. During deposition, the co-sputtered Ta$_2$O$_5$ and ZrO$_2$ particles impinged on substrates heated up to about 120 $^\circ$C.

Single layers with a thickness between 500 and 650 nm have been grown on various substrates for different purposes: (i) on fused-silica witness samples (50 mm in diameter, 6-mm thick) for transmission spectrophotometry; (ii)  on silicon wafers (75 mm in diameter, 0.5-mm thick) for reflection spectroscopic ellipsometry and EDX analyses; (iii) and on a Corning 7980 fused-silica disk-shaped mechanical resonator of 75 mm in diameter and 1-mm thick for measuring coating mechanical loss, Young modulus, Poisson ratio and density. Additionally, Ta$_2$O$_5$-ZrO$_2$/SiO2 HR coatings with the same design of those of Advanced LIGO and Advanced Virgo \cite{Granata20} have been deposited on micro-polished fused-silica witness samples for optical absorption and scattering measurements.

Prior to coating deposition, to release the internal stress due to manufacturing and induce relaxation, the fused-silica disk was annealed in air at 900 $^\circ$C for 10 hours. After deposition, all coating samples were annealed in air for 10 hours: optical samples were annealed at 500 $^\circ$C, the disk underwent consecutive annealing treatments of increasing peak temperature (500, 600 and 700 $^\circ$C) and the mechanical loss of its coating was measured after each annealing step.

We used an analytical balance to measure the mass of the disk before and after the coating deposition, as well as after the 500-$^\circ$C annealing. To estimate the surface area coated in the deposition process, we measured its diameter with a Vernier caliper. As the coating thickness was known with high accuracy from the optical characterization (see Section \ref{SECT_LMAopt}), the coating density $\rho = 6.9 \pm 0.1$ g/cm$^3$ was straightforwardly estimated as the mass-to-volume ratio.

We used a Zeiss LEO 1525 field-emission scanning electron microscope (SEM) and a Bruker Quantax system equipped with a Peltier-cooled XFlash 410-M silicon drift detector to analyze the surface and elemental composition of the single-layer coatings. The SEM beam was set to 15 keV for the survey. We performed multiple energy-dispersive X-ray (EDX) analyses on different coating sample spots and with different magnifications (from $100\times$ to $5000\times$). Semi-quantitative (standardless) results were based on a peak-to-background evaluation method of atomic number, self-absorption and fluorescence effects (P/B-ZAF correction) and a series fit deconvolution model provided by the Bruker Esprit 1.9 software. Using the self-calibrating P/B-ZAF standard-based analysis, no system calibration had to be performed. According to our EDX analyses, our thin films had an average atomic cation ratio $\eta =$ Zr/(Zr+Ta) = 19 $\pm $ 1 \%. We also found that they incorporated 2.9 $\pm $ 0.6 at.\% of Ar.

\subsection{ECR-IBD}
\label{SECT_samplesECR}
Coatings were deposited using a custom-built ion beam deposition system that utilized three ECR ion sources for sputtering~\cite{Birney2019}. The ion beams are generated by injection of Ar gas into a resonance microwave cavity, with $\sim$3 W injected power at 2.45 GHz. The confinement of the plasma permits the extraction of ions through a single aperture, enabling a larger range of stable extraction potentials to be explored, with low ion current. The extraction potential was 10 kV and ion currents ranged between  $0.2 - 0.6$ mA per source, resulting in a deposition rate of 0.001 nm/s. Base pressure was below $1 \times 10^{-5}$ mbar prior to deposition, with pressure rising to $8 \times 10^{-5}$ mbar with Ar injected through the ion sources, and increasing to $1.2 \times 10^{-4}$ mbar with reactive O$_2$ gas fed into the chamber. The targets used during the deposition process are two circular 152.4-mm diameter targets of pure tantala and zirconia provided by Scotech, UK, both with $99.99\%$ purity. During the deposition process, two ion beams were directed onto the tantala target and one beam on the zirconia target, with the aim for cation ratio $\eta =$ Zr/(Zr+Ta) of $\sim$33\%. Fused-silica witness samples and cantilevers were placed on a rotating substrate holder to ensure coating uniformity.

The cantilevers, used for coating mechanical loss measurements \cite{Vajente_2018}, were fabricated from 45 mm $\times$ 5 mm $\times$ $(175\pm3)$ $\mu$m strips of fused silica, flame welded to 10 mm $\times$ 10 mm $\times$ 3.1 mm clamping blocks \cite{Vajente_2018}. They were cleaned in boiling H$_2$O$_2$ (30\% w/w in H$_2$O) with 1 mol/l KOH dissolved, in order to remove the majority of particulates associated with the welding process, then annealed in air for 5 hours at 950 $^{\circ}$C prior to measurement and deposition.

EDX was used to quantify the cation ratio $\eta =$ Zr/(Zr+Ta), found to be $34.0 \pm 0.5$\%.

\subsection{RBTD}
Coatings were deposited using in a LANS system by 4Wave, Inc \cite{zhurin2000biased}, where a low energy Ar ion plume is produced using an End-Hall ion source operated at 60 V and 5.8 A, with an Ar flow of 27 sccm. For neutralization, a hollow cathode electron source is employed at an emission current of 6.2 A with an Ar flow of 25 sccm. The Ar ions are attracted to a biased metallic target, leading to sputtering of the target material. For oxide coatings, O$_2$ is introduced into the chamber during the process. Substrates are placed in a sample holder facing the ion source, and rotating at 20 rpm during deposition to ensure uniformity of the film. The chamber base pressure is below $1 \times 10^{-7}$ mbar, while the process pressure is about $9 \times 10^{-4}$ mbar.

The LANS system has the ability to simultaneously bias up to three targets to produce mixed oxide coatings. An asymmetric, bipolar DC pulse is used to bias the targets with a negative voltage of -800 V and a positive voltage of 10 V and a period of 100 $\mu$s. For a binary oxide mixture, two targets are simultaneously biased and different mixture proportions of the materials are achieved by controlling the individual pulse duration of each target. Ta and Zr targets were employed to deposit tantala-zirconia thin films, and the corresponding deposition conditions are presented in Table \ref{TAB_deposParams_RBTD}. Deposition rates for RBTD are typically an order of magnitude lower than with a gridded IBS system, ranging from 0.0212 nm/s for tantala to 0.0116 nm/s for zirconia. Films with a thickness between 200 - 300 nm were grown on: (i) 25.4-mm diameter and 6.35-mm thick UV grade fused silica substrates; (ii) 75-mm diameter and 1-mm thick fused silica substrates; and (iii) (111) Si wafer substrates.
\begin{table*}[!ht]
	\setlength\tabcolsep{5pt}
	\begin{center}
	\begin{tabular}{ccccc}
		\hline
		Film & O$_2$ flow [sccm] & Target & Pulse width [$\mu$s] & Deposition rate [nm/s]\\
		\hline
		CSU I & 12 & Ta & 2 & 0.0212  $\pm$ 0.0001\\
		CSU II & 12 & Ta - Zr & 2 - 54 & 0.0294  $\pm$ 0.0001\\
		CSU III & 12 & Ta - Zr & 47 - 2 &  0.0217 $\pm$ 0.0001\\
		CSU IV & 12 & Zr & 2 & 0.0116  $\pm$ 0.0001\\
		\hline
	\end{tabular}
	\end{center}
	\caption{Deposition conditions for Ta$_2$O$_5$-ZrO$_2$ thin films grown via RBTD. Pure Ta$_2$O$_5$ and ZrO$_2$ samples (I and IV, respectively) are included for reference.}
	\label{TAB_deposParams_RBTD}
\end{table*}

X-ray photoelectron spectroscopy (XPS) measurements were performed on  as-deposited coatings using a Physical Electronics PE 5800 ESCA/ASE system with a monochromatic Al K$\alpha$ x-ray source. The photoelectron take-off angle was set at $45^{\circ}$ and a charge neutralizer with a current of 10 $\mu$A was employed. The chamber base pressure was around $1 \times 10^{-9}$ mbar. High resolution scans were recorded for the main peaks of the detected elements: C 1s, O 1s, Zr 3d and Ta 4f. No Ar can be detected by XPS in this case, because the peak associated with this element is superimposed with a Ta peak. Presence of carbon is due to normal surface contamination of  samples exposed to the atmosphere. In the case of insulating samples the carbon is not removed but instead used for calibration of the binding energy scale, by employing the position of the adventitious carbon peak. CasaXPS software (version 2.3.19) was used to fit the high-resolution spectra \cite{casaxps}.
\begin{table*}[!ht]
	\begin{center}
		\begin{tabular}{cccccc}
			\hline
			\multirow{2}{*}{Film} & \multicolumn{4}{c}{Elemental concentration [at.\%]} & Cation ratio \\
			\cline{2-5}
			& C & O & Zr & Ta & Zr/(Zr+Ta)\\
			\hline
			CSU II & 38 $\pm$ 1 & 44 $\pm$ 1 & 4.3 $\pm$ 0.1 & 14.1 $\pm$ 0.1 & 0.23 $\pm$ 0.01\\
			CSU III & 39 $\pm$ 1 & 42 $\pm$ 1 & 10.4 $\pm$ 0.4 & 8.7 $\pm$ 0.1 & 0.54 $\pm$ 0.03\\
			\hline
		\end{tabular}
	\end{center}
	\caption{Atomic concentrations obtained from XPS analysis at the surface of as-deposited Ta$_2$O$_5$-ZrO$_2$ thin films grown via RBTD.}
	\label{TAB_xps_RBTD}
\end{table*}

Table \ref{TAB_xps_RBTD} shows the atomic concentrations obtained from XPS for the thin films as deposited. The atomic percentage cation ratio, $\eta =$ Zr/(Zr+Ta), resulted in 0.23 $\pm$ 0.01 for sample II and 0.54 $\pm$ 0.03 for sample III.
\begin{figure}[h!]
	\centering
	\subfloat[][]{
		\includegraphics[width=0.40\linewidth]{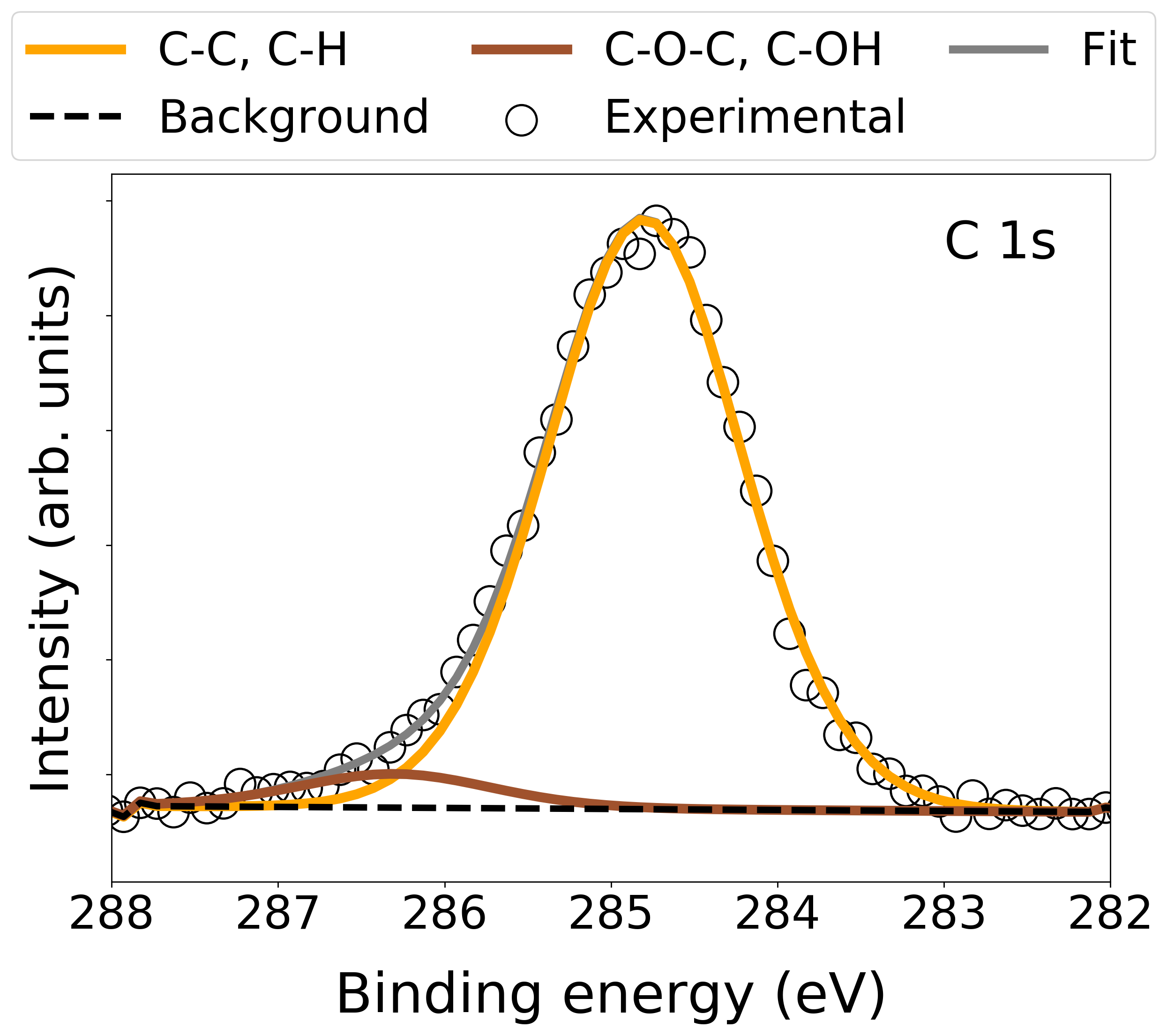}}
	\subfloat[][]{
		\includegraphics[width=0.40\linewidth]{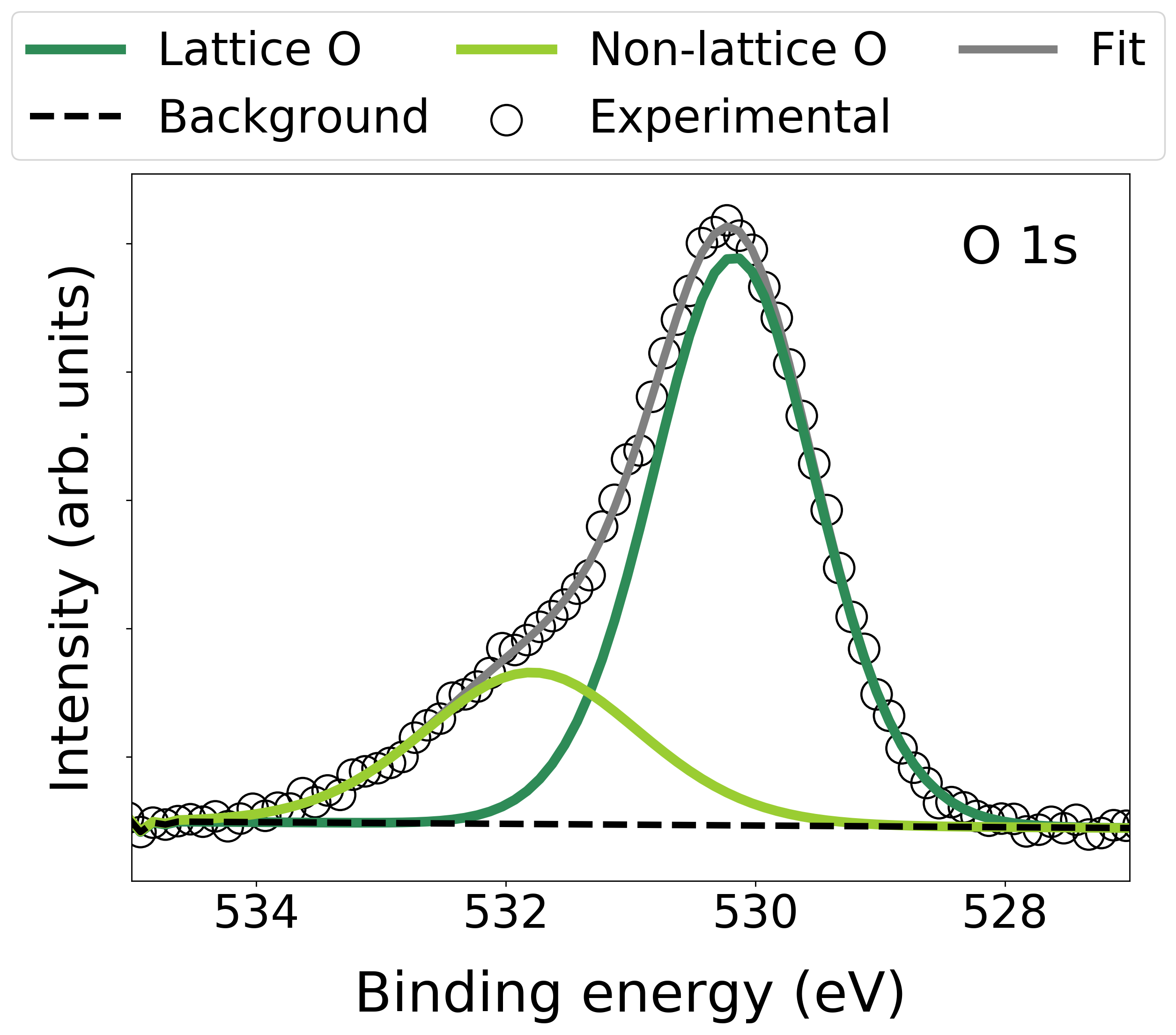}}\\
	\vspace{0.2cm}
	\subfloat[][]{
		\includegraphics[width=0.40\linewidth]{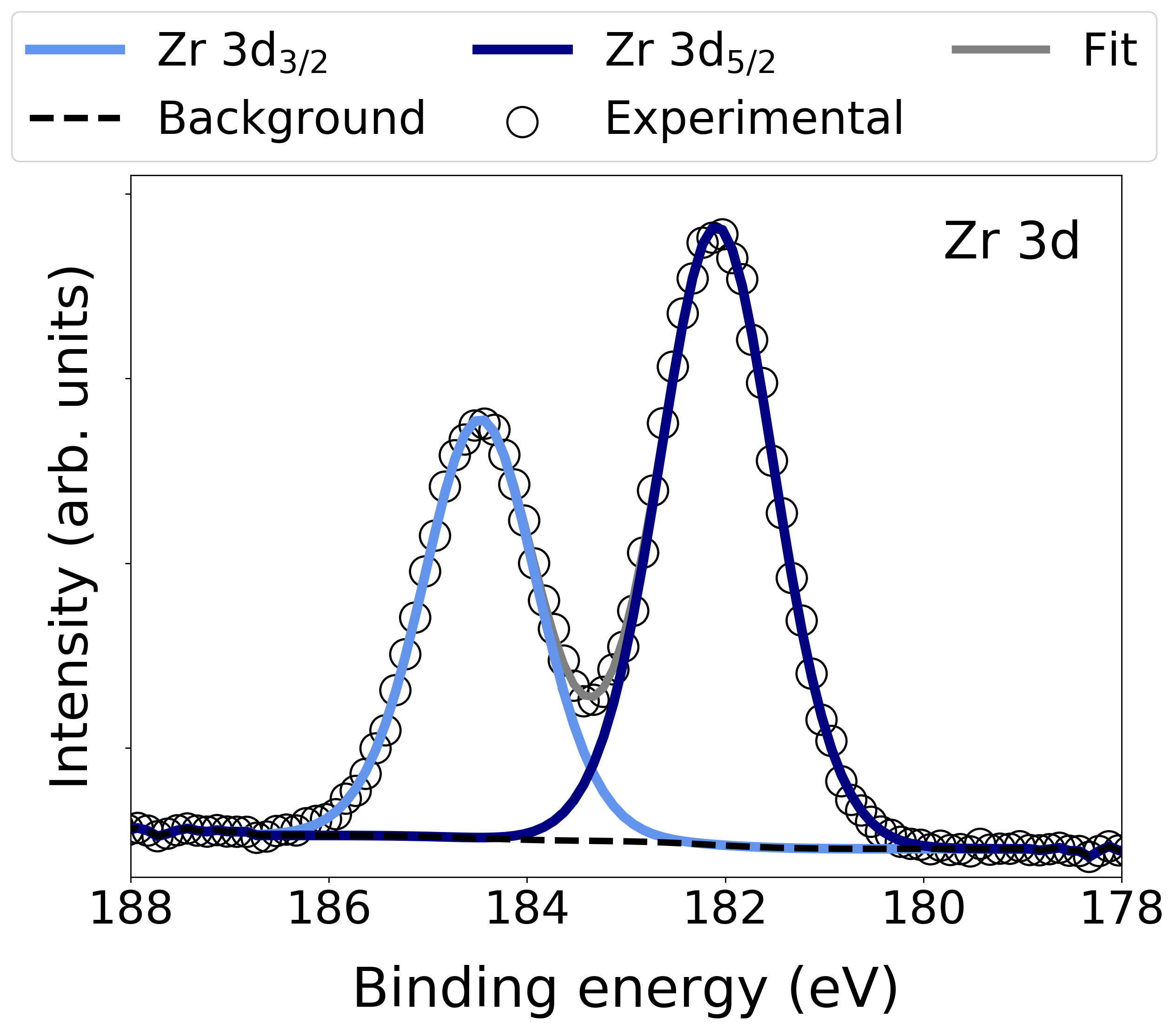}}
	\subfloat[][]{
		\includegraphics[width=0.40\linewidth]{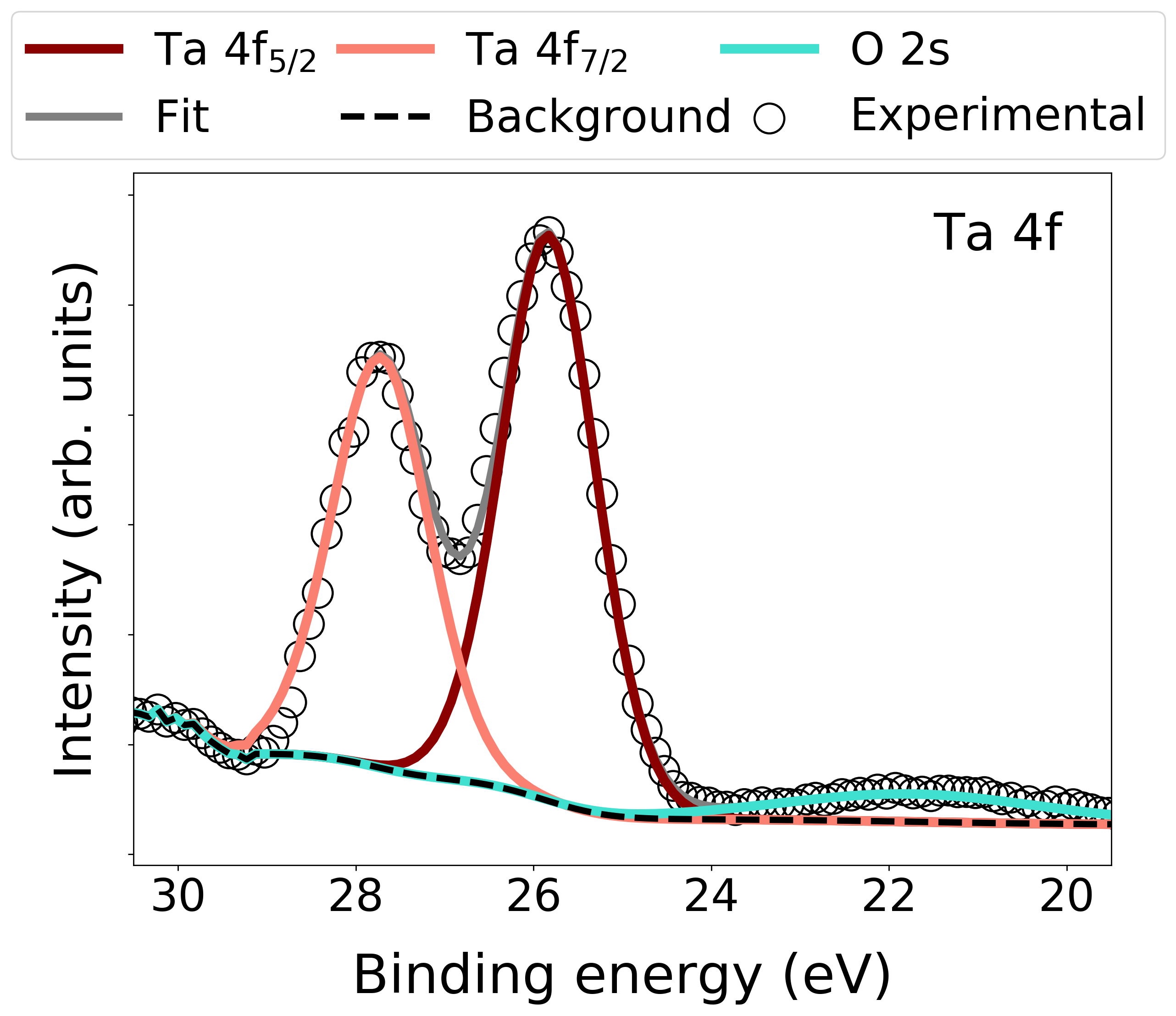}}
	\caption{\label{fig:xps_csu} Typical fit of XPS peaks for an as-deposited Ta$_2$O$_5$-ZrO$_2$ thin film grown via RBTD with a cation ratio of 0.54: (a) C 1s, (b) O 1s, (c) Zr 3d and (d) Ta 4f.}
\end{figure}

Fig. \ref{fig:xps_csu} presents typical fits of the main XPS peaks. The C 1s peak was composed mainly of adventitious carbon, and up to 5\% of carbon was found to be linked to an alcohol and/or esther functionality. This further confirmed that the presence of carbon was due to surface contamination. The O 1s peak had two components: lattice O, determined as such based on the tabulated position of O 1s for zirconia and tantala (around 530 eV), and non-lattice oxygen. In all cases, 1 to 3\% of the oxygen was associated with hydroxide species based on the analysis of the C 1s peak, so the non-lattice peak could be attributed mainly to oxygen in defective sites. In the case of the Zr 3d peak, the doublet energy separation was found to be (2.39 $\pm$ 0.01) eV and the Zr 3d$_{5/2}$ peak position was (182.1 $\pm$ 0.1) eV for both films. These values are in agreement with the tabulated values for zirconia \cite{nist_database}. For the Ta 4f doublet in both films, the energy separation was determined as (1.89 $\pm$ 0.01) eV and the Ta 4f$_{7/2}$ peak position as (26.0 $\pm$ 0.2) eV, which is in agreement with reported values for tantala \cite{nist_database}. Therefore, for the evaluated zirconia-doped tantala films as deposited, there was no noticeable modification of the chemical environment of either Ta or Zr, compared to their corresponding oxides.

\subsection{HiPIMS/RF-MS}
Layers of approximately 700 nm in thickness were synthesized in a Kurt J. Lesker CMS-18 system, equipped with two 76-mm diameter metallic targets of Zr and Ta of 99.95\% purity located $\sim$25 cm beneath the substrate holder. Prior to deposition, the base pressure in the  system was less than  $10^{-7}$ mbar. Depositions were carried out on different samples: (i) the $75$-mm diameter fused-silica disks were placed face down, one at the time, on a rotating substrate holder featuring a wedged hole in the middle, such that they were supported by a $1$ mm ledge; (ii) silica and (iii) silicon substrates were placed next to the disks as witness samples. During deposition, the pressure was 9 mbar with a 20\% O2:Ar gas ratio.

For six samples, Ta was sputtered by HiPIMS \cite{Hala2014a, Hala2014b} using a Huttinger power supply, while the Zr was sputtered using a  power supply (13.56 MHz) \cite{BRAUER20101354}. Two additional samples had their Ta deposited by RF-MS while the Zr was deposited by DC-MS, for comparison. The nominal duty cycle of the HiPIMS process was set at a voltage of 700 V, with a pulse duration of $80$ Hz. The nominal RF voltage applied to the Zr target varied between $200$ V and $400$ V, in order to achieve different doping levels. Both magnetrons were pre-sputtered for 12 minutes prior to deposition, in order to reach steady-state conditions before opening the main shutter. Depositions typically lasted 2 to 6 hours.

Rutherford back scattering (RBS) measurements \cite{chu2012backscattering} were performed on the Si witness samples to measure the composition and areal atomic density of the films. Two ion beam configurations were implemented: a 2 MeV He ion beam to obtain the classical Rutherford cross section of O, and a 3.9 MeV He ion beam in order to separate the different peaks of Ta, Zr and Ar. The latter allowed for a higher precision of the concentration of these elements but with less precision on the O content, given the non-Rutherford cross-section and charging effects. The incidence angle was set at $7^{\circ}$ and the scattering angle at $170^{\circ}$ to maximize mass resolution. SIMNRA simulations \cite{MayerRBS} were compared to the experimental spectra in order to extract the composition and the areal atomic density of the deposited layers. Uncertainties were estimated by varying the concentrations until the simulation no longer fit the data. The resulting cation ratios $\eta =$ Zr/(Ta+Zr) are reported in Table \ref{TAB_samples}, while the O content is discussed below. The coatings contain $\sim$1.5\% Ar, as expected for samples deposited in Ar ambient. Elastic recoil detection analysis in other coatings usually also reveals the presence of 1.5 to 3.0\% of hydrogen. We expect this to be the case in these films as well. No concentration gradients were identified within the sensitivity of our measurements.

The mass density of the films was calculated by the following two methods. The first one relies on the areal atomic density (at./cm$^2$) and composition found by RBS. Knowing the atomic mass, it was possible to compute the areal mass density (g/cm$^2$). The second method consisted in measuring the substrate mass before and after deposition using a micro-balance, and dividing it by the coated area. The latter was measured from a photograph, in order to account for the area of the uncoated region near the edges of the sample. For both methods, the areal density was divided by the thickness found by ellipsometry (see Section \ref{SECT_MS_optical}). We note that for the density calculation based on RBS, due to the presence of thickness gradients, both measurements needed to be carried out at the same location to avoid discrepancies. Likewise, in the case of the weighing method, it was important to consider an average thickness deduced from the thickness mapping obtained by ellipsometry.
\begin{figure}
\centering
\includegraphics[width=0.40\linewidth]{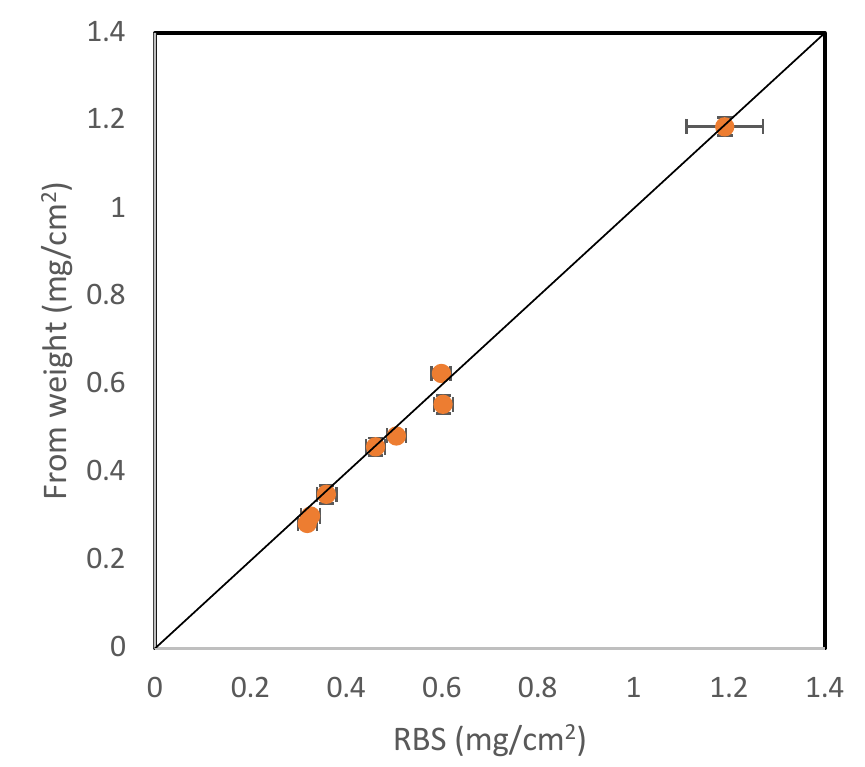}\hfill
\caption{Areal mass density of Ta$_2$O$_5$-ZrO$_2$ thin films grown via MS: values measured by sample weighing are presented as a function of those measured by RBS, for comparison. The black line shows the 1:1 relationship.}\label{FIG_MS_RBSvsWEIGHT}
\end{figure} 
Figure~\ref{FIG_MS_RBSvsWEIGHT} compares the areal mass density deduced from RBS measurements and by weighing the samples, that is, without considering the thickness. We can observe that both data sets closely follow a 1:1 relationship. Given the independence of both methods, we can assume that the measured values are accurate. Densities reported in Table \ref{TAB_samples} are average values between the two methods.
%
\section{Optical properties}
\label{SECT_opt}
The optical properties of the coating samples were characterized using different techniques. Table \ref{TAB_optical_results} presents the measured values for the refractive index $n$ at the wavelengths of interest (1064 and 1550 nm) for current and future GW interferometers \cite{aLIGO, aVirgo, einstein-telescope, cosmic-explorer}. Optical absorption at 1064 nm was measured on RBTD single-layer samples and IBS multi-layer HR stacks. Light scattering at 1064 nm was measured on IBS multi-layer HR stacks. At 1550 nm, where only ellipsometric and photometric measurements were available, extinction values were smaller than the sensitivity of the instruments ($k < 10^{-3}$). Measurements and their analyses are described below. A comparative summary is presented in Section \ref{SECT_opt_compar}, where correlations between oxygen concentration, density and refractive index are discussed.
\begin{table*}[!ht]
	\begin{center}
		\begin{tabular}{ccccccccc}
			\hline
			Sample & Process & $\eta$ [at.\%] & $n_{1064}$ & $n_{1550}$\\
			\hline
            MLD2014 & IBS & $48.5 \pm 0.4$\\
            MLD2018 & IBS & $50.2 \pm 0.6$ & $2.08\pm0.01$\\
            LMA & IBS & 19 $\pm$ 1 & 2.09 $\pm$ 0.01 & 2.07 $\pm$ 0.01\\
            UoS & ECR-IBD & $34.0 \pm 0.5$ & 1.96 $\pm$ 0.01 & 1.95 $\pm$ 0.01\\
            UMP 551 & RF-MS & 41 $\pm$ 3 & 2.01 $\pm$ 0.01 & 1.99 $\pm$ 0.01\\ 
            UMP 554 & RF-MS & 43 $\pm$ 3 & 1.99 $\pm$ 0.01 &1.97 $\pm$ 0.01\\ 
            UMP 658 & HiPIMS & 24 $\pm$ 3 & 2.13 $\pm$ 0.01 & 2.11 $\pm$ 0.01 &\\ 
            UMP 659 & HiPIMS & 24 $\pm$ 2 & 2.11 $\pm$ 0.01 & 2.10 $\pm$ 0.01 &\\ 
            UMP 542 & HiPIMS & 7 $\pm$ 2 & 2.12 $\pm$ 0.01 & 2.11 $\pm$ 0.01 &\\ 
            UMP 664 & HiPIMS & 7 $\pm$ 2 & 2.11 $\pm$ 0.01 & 2.10 $\pm$ 0.01 &\\ 
            UMP 678 & HiPIMS & 47 $\pm$ 2 & 2.11 $\pm$ 0.01 & 2.10 $\pm$ 0.01 &\\ 
            UMP 680 & HiPIMS & 47 $\pm$ 2 & 2.11 $\pm$ 0.01 & 2.10 $\pm$ 0.01 &\\ 
            CSU II & RBTD & 23 $\pm$ 1 & 2.07 $\pm$ 0.01 & 2.06 $\pm$ 0.01\\
            CSU III & RBTD & 54 $\pm$ 3 & 2.10 $\pm$ 0.01 & 2.09 $\pm$ 0.01\\
			\hline
		\end{tabular}
	\end{center}
	\caption{Measured refractive index $n$ at 1064 and 1550 nm of co-sputtered Ta$_2$O$_5$-ZrO$_2$ thin films. RF-MS samples had Ta deposited by RF-MS and Zr deposited by DC-MS, HiPIMS samples had Ta deposited by HiPIMS and Zr deposited by RF-MS. Index values are plotted and discussed in Section \ref{SECT_opt_compar}.}
	\label{TAB_optical_results}
\end{table*}

\subsection{IBS}
\subsubsection{MLD --}
\begin{centering}
\begin{figure}[h!]
\hfil\subfloat[][]{\includegraphics[width = 0.40\textwidth]{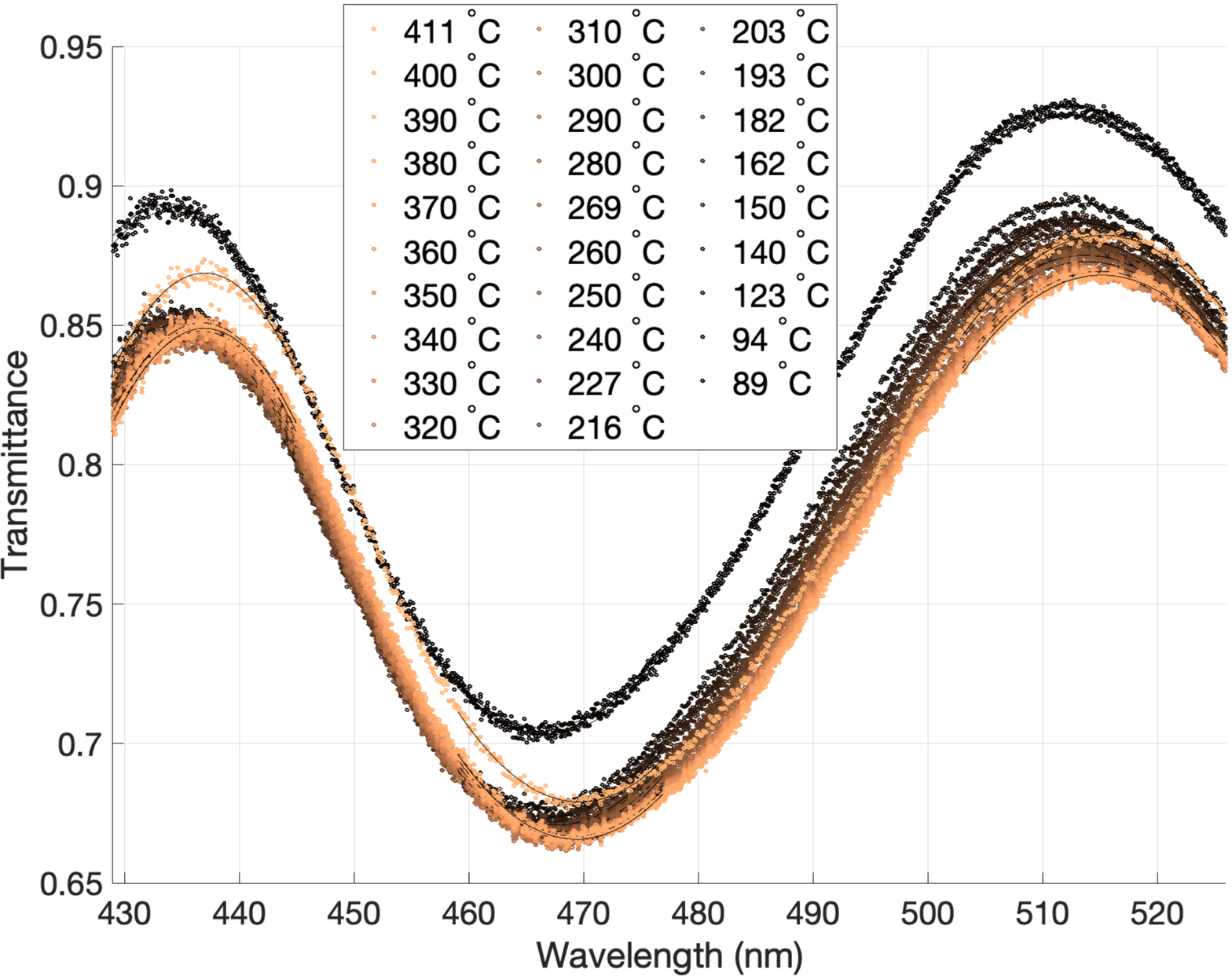} \label{FIG_MLD_extrema_fits}}\hfil
\hfil\subfloat[][]{\includegraphics[width = 0.40\textwidth]{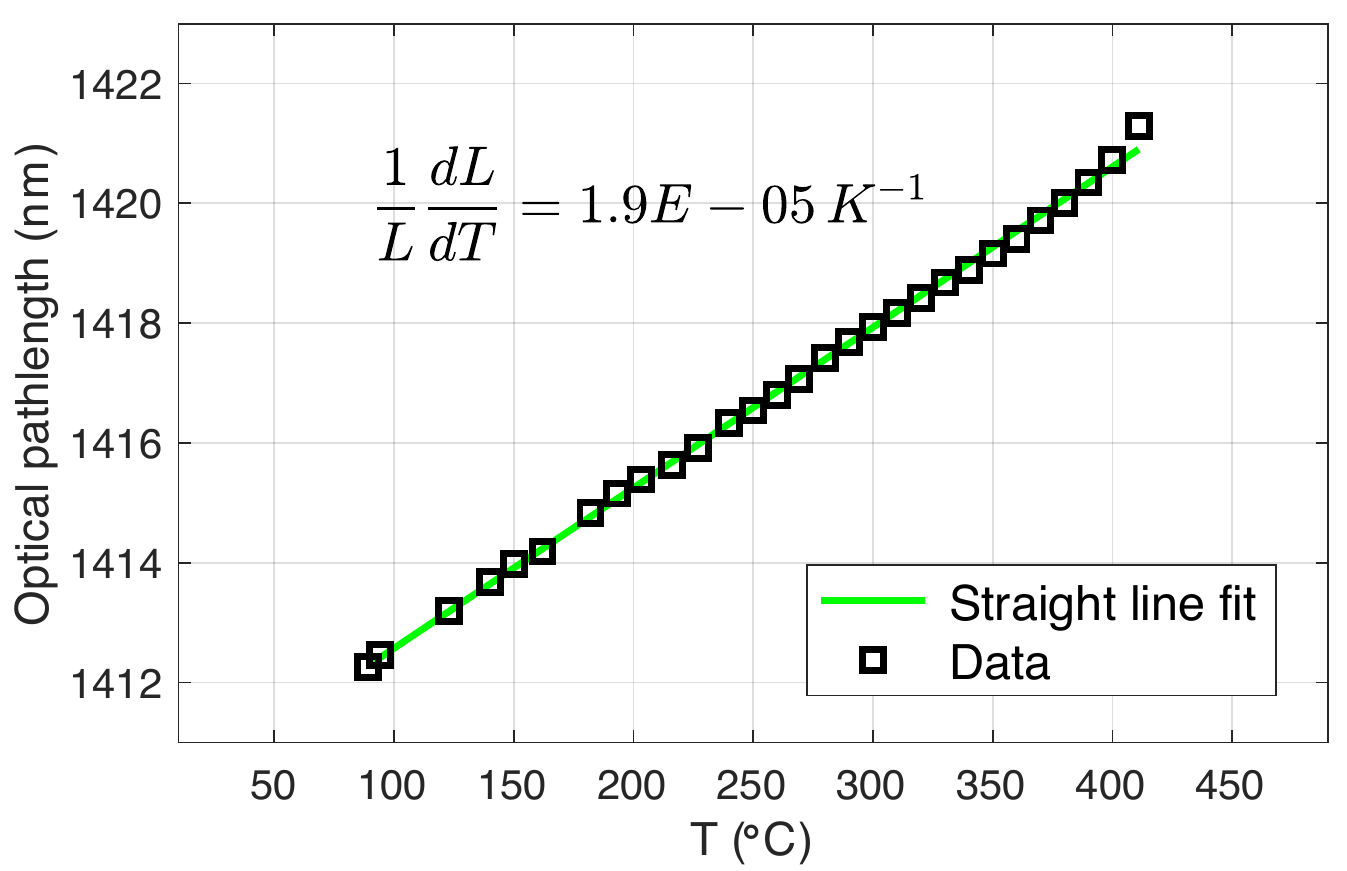} \label{FIG_MLD_optical_path_vs_T}}\hfil
\caption{Characterization of optical thickness dependence on temperature of IBS Ta$_2$O$_5$-ZrO$_2$ thin film MLD2014. (a) Transmission spectrum extrema with parabolic fits to determine the extrema locations as a function of temperature. Higher temperatures shifts the extrema to longer wavelengths, indicating an increasing optical thickness with temperature. (b) Optical thickness of the coating as a function of temperature.}
\end{figure}
\end{centering}

The manufacturer reported a refractive index of $n=2.08 \pm 0.01$ at 1064 nm for sample MLD2018. We measured the optical thickness dependence on temperature for sample MLD2014 by monitoring the wavelength dependence of its transmission spectrum extrema as a function of temperature \cite{Gretarsson_dndT}, in the range $\mathrm{89-411}$ $^\circ$C. Typical fits to the extrema locations are shown in Figure~\ref{FIG_MLD_extrema_fits}. The measured coating optical thickness as a function of temperature, shown in Figure~\ref{FIG_MLD_optical_path_vs_T}, is linear. Note that, due to uncertainty in the physical thickness of the coating layer, optical thickness is not used here to estimate the absolute value of the index, only the change with temperature. The fractional optical thickness change as a function of temperature was found to be
\begin{equation}
\frac{1}{L}\frac{\mathrm{d}L}{\mathrm{d}T} = \alpha + \beta/n = (18.6\pm1)\times10^{-6}\,\mathrm{K^{-1}}\ ,
\end{equation}
where $\alpha$ is the thermal expansion coefficient and $\beta$ is the thermo-optic coefficient.

\subsubsection{LMA --}
\label{SECT_LMAopt}
Refractive index and thickness of as-deposited coating samples were determined by spectroscopic ellipsometry, via coupled analysis of data sets acquired with two different J. A. Woollam commercial instruments covering complementary spectral regions: a VASE for the 190-1100 nm range and a rotating-compensator M-2000 for the 245-1700 nm range. The wide range swept with both instruments allowed us to extend the analysis from ultraviolet to near infrared (0.7 - 6.5 eV). Consistent results were obtained in the overlapping range, within the uncertainty of the models used to fit the data. Analysis was carried out independently for VASE and M-2000 data, and eventually the information was combined in order to obtain the final results.

The optical constants were obtained by measuring the amplitude ratio $\Psi$ and phase difference $\Delta$ of the p- and s-polarized light \cite{Fujiwara07} reflected by the samples at different incidence angles $\theta = 55$\textdegree, $60$\textdegree, $65$\textdegree, close to the coating Brewster angle. The optical response of the substrates (silicon wafers with a thin native oxide layer and unpolished rear surface in order to avoid backside spurious reflections) was characterized with prior dedicated measurements and analyses \cite{Amato19a}. 
As in previous works on similar systems \cite{Amato19a,prato2011gravitational}, we analyzed the data through a three-layer model consisting of substrate, thin film and surface layer. The latter, accounting for surface roughness, was defined via a Bruggeman effective medium approximation. The resulting coating roughness was less than 1 nm, a value comparable with those measured with an atomic-force microscope on similar samples \cite{prato2011gravitational}. Being less than 1\% of the coating thickness, the surface layer did not affect substantially the data fit, especially in the near-infrared region.

The dielectric functions were reconstructed by using Kramers-Kronig-consistent Tauc-Lorentz \cite{Jellison1996} and Cody-Lorentz \cite{ferlauto2002analytical} models. Both models fit the data with comparable mean squared error, and yielded consistent results. We then obtained the complex refractive index of the as-deposited thin films from the best-fit dielectric functions. The Cody-Lorentz model provided a slightly better fit in the region of the absorption edge and included also the Urbach tail absorption, which may be correlated to coating mechanical loss \cite{Amato20}. However, data quality for energies higher than the energy gap ($E > 4$ eV) was degraded and resulted in a large uncertainty for the Urbach energy $E_U = 130 \pm 30$ eV.

\begin{figure}
\centering
    \includegraphics[height=0.35\textwidth]{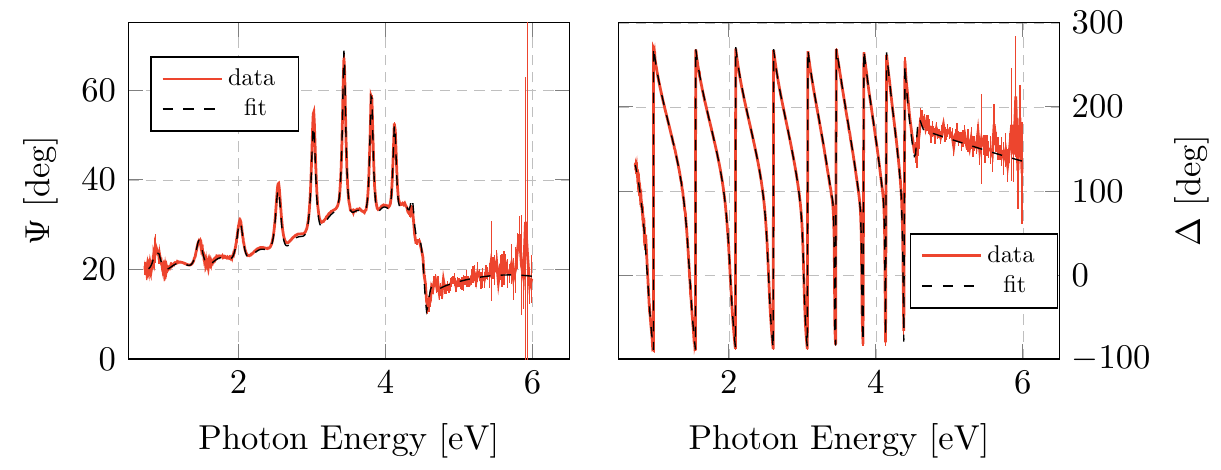}
	\caption{Measured ellipsometric spectra of an as-deposited IBS Ta$_2$O$_5$-ZrO$_2$ thin film (LMA), acquired at an incidence angle $\theta$ = 60$^{\circ}$. The best-fit Cody-Lorentz model is shown (dashed line).}
	\label{FIG_LMA_psidelta}
\end{figure}

\begin{figure}[h!]
	\centering
	\subfloat[]
	{\includegraphics[height=0.35\textwidth]{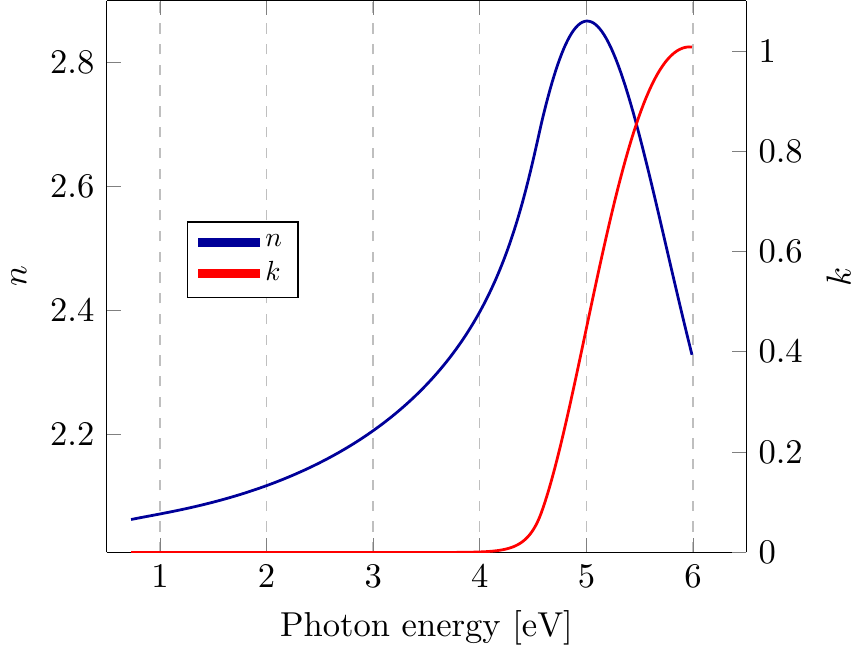}\label{FIG_LMA_n-k}}\quad
	\subfloat[]
	{\includegraphics[height=0.35\textwidth]{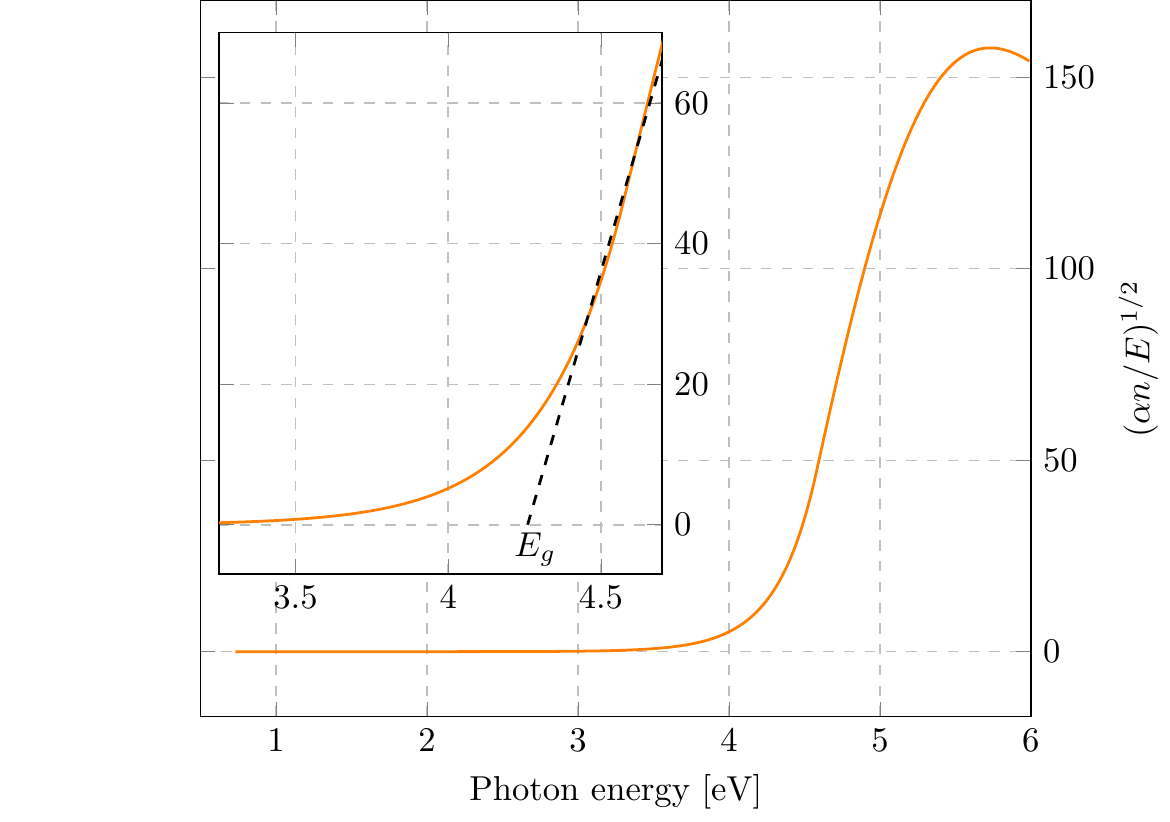}\label{FIG_LMA_cody}}
	\caption{Measured optical properties of as-deposited IBS Ta$_2$O$_5$-ZrO$_2$ thin films (LMA): (a) refractive index $n$ and extinction coefficient $k$ as function of photon energy $E$, the region of interest for present GW detectors is around $E=1.2$ eV (corresponding to a wavelength of 1064 nm); (b) Cody plot highlighting the energy gap $E_g$.}
	\label{FIG_LMA_optics}
\end{figure}

All measured spectra were fit with the same accuracy. Figure~\ref{FIG_LMA_psidelta} shows exemplary spectra acquired at $\theta = 55^{\circ}$ and their best fits with a Cody-Lorentz model. Figure~\ref{FIG_LMA_n-k} shows the reconstructed coating optical constants, and Figure~\ref{FIG_LMA_cody} is a Cody plot highlighting the coating energy gap, $E_g = 4.1 \pm 0.1$ eV, which is consistent with that of pure tantala coatings produced under strictly identical conditions \cite{Amato19a}.

After annealing at 500~$^{\circ}$C in air for 10 hours, single-layer coating samples were characterized in their transparent region (400-1400 nm) with a Perkin Elmer Lambda 1050 spectrophotometer, at normal incidence. The annealing reduced the refractive index in the near-infrared region by about 3\%. Samples of HR stacks, also annealed at 500 $^{\circ}$C in air for 10 hours, were characterized for optical absorption and light scattering with a custom-developed setup based on the photo-thermal deflection principle \cite{Boccara80} and a commercial CASI scatterometer, respectively. We measured an optical absorption of 0.5 parts per million (ppm), and 45 ppm of scattered light.

\subsection{ECR-IBD}
\begin{figure}[h!]
	\centering
		\includegraphics[width=0.55\linewidth]{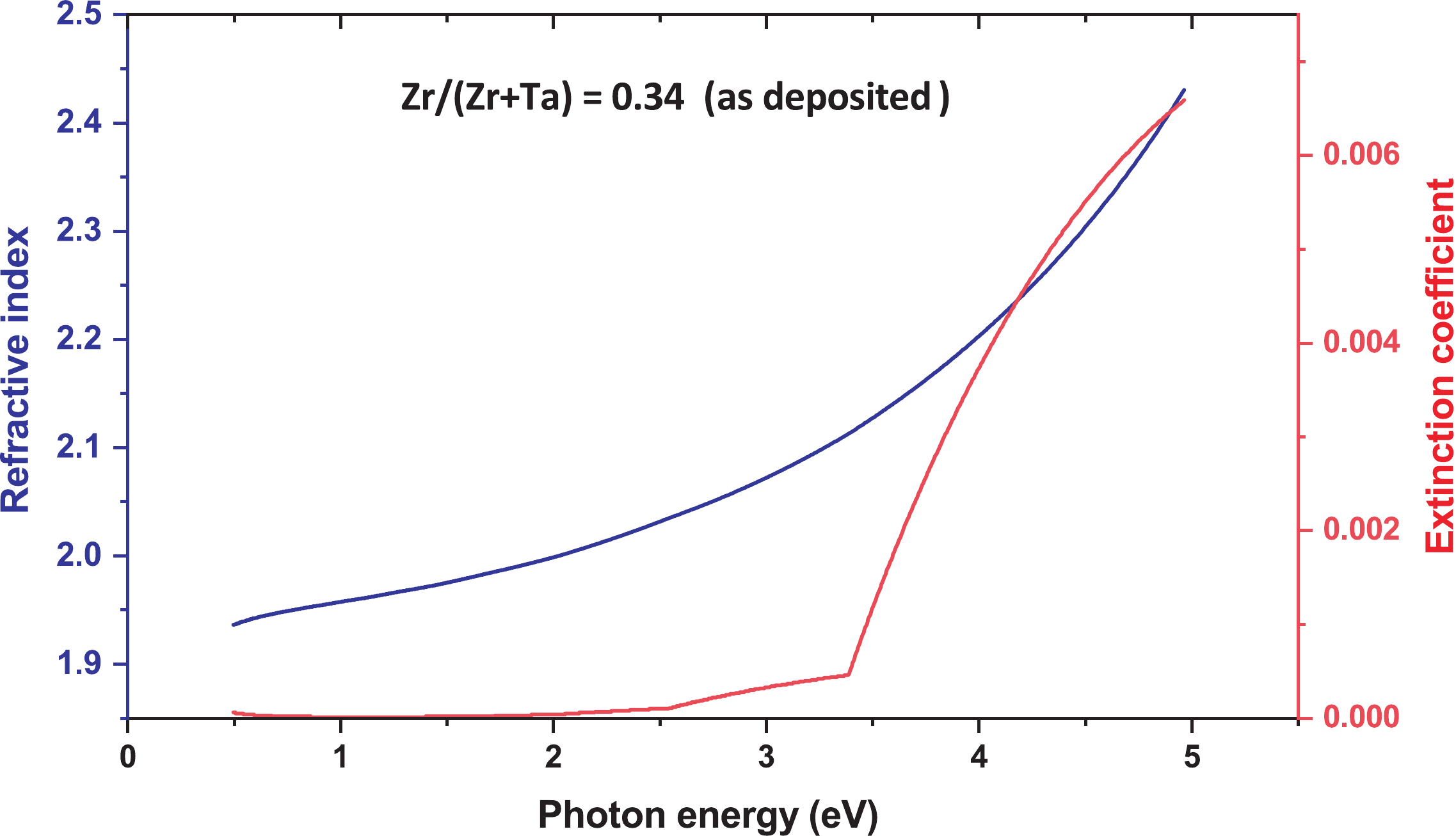}
	\caption{\label{FIG_ECR_n-k} Refractive index $n$ and extinction coefficient $k$ as function of photon energy for as-deposited ECR-IBD Ta$_2$O$_5$-ZrO$_2$ thin films with cation ratio of 0.34.}
\end{figure}

Optical-transmission spectra of as-deposited films, shown in Figure~\ref{FIG_ECR_n-k}, were measured in the 250-2100 nm range using a Varian 5000 spectrophotometer and fit to transmission simulations using SCOUT software by WTheiss. At 1064 nm and 1550 nm, the refractive index $n$ was found to be $1.96 \pm 0.01$ and $1.95 \pm 0.01$, respectively, and the extinction coefficient $k < 10^{-3}$.

\subsection{RBTD}
\label{SECT_RBTDopt}
Refractive index and extinction coefficient for the films were determined by spectroscopic ellipsometry using a Horiba UVISEL ellipsometer. Measurements were taken at an angle of $60^{\circ}$, for a spectral range of 0.59 - 6.5 eV. The films were evaluated as deposited and after annealing at the maximum temperature for which no crystallization was detected and mechanical loss minimized. A reduction of the refractive index was found after annealing, which could be due to oxygen incorporation during the process or due to formation of Ar bubbles \cite{fazio2020structure}. Films previously grown by our group were found to contain 3 to 5 \% of Ar. When Ar bubbles are formed and coalesce upon annealing, the films usually feature a reduction in the refractive index that is accompanied by an increase in thickness. In our case, however, the thickness did not increase after annealing within the resolution of the instrument, indicating no significant presence of Ar bubbles in the coatings.

\begin{figure}[h!]
	\centering
	\subfloat[][]{
		\includegraphics[width=0.45\linewidth]{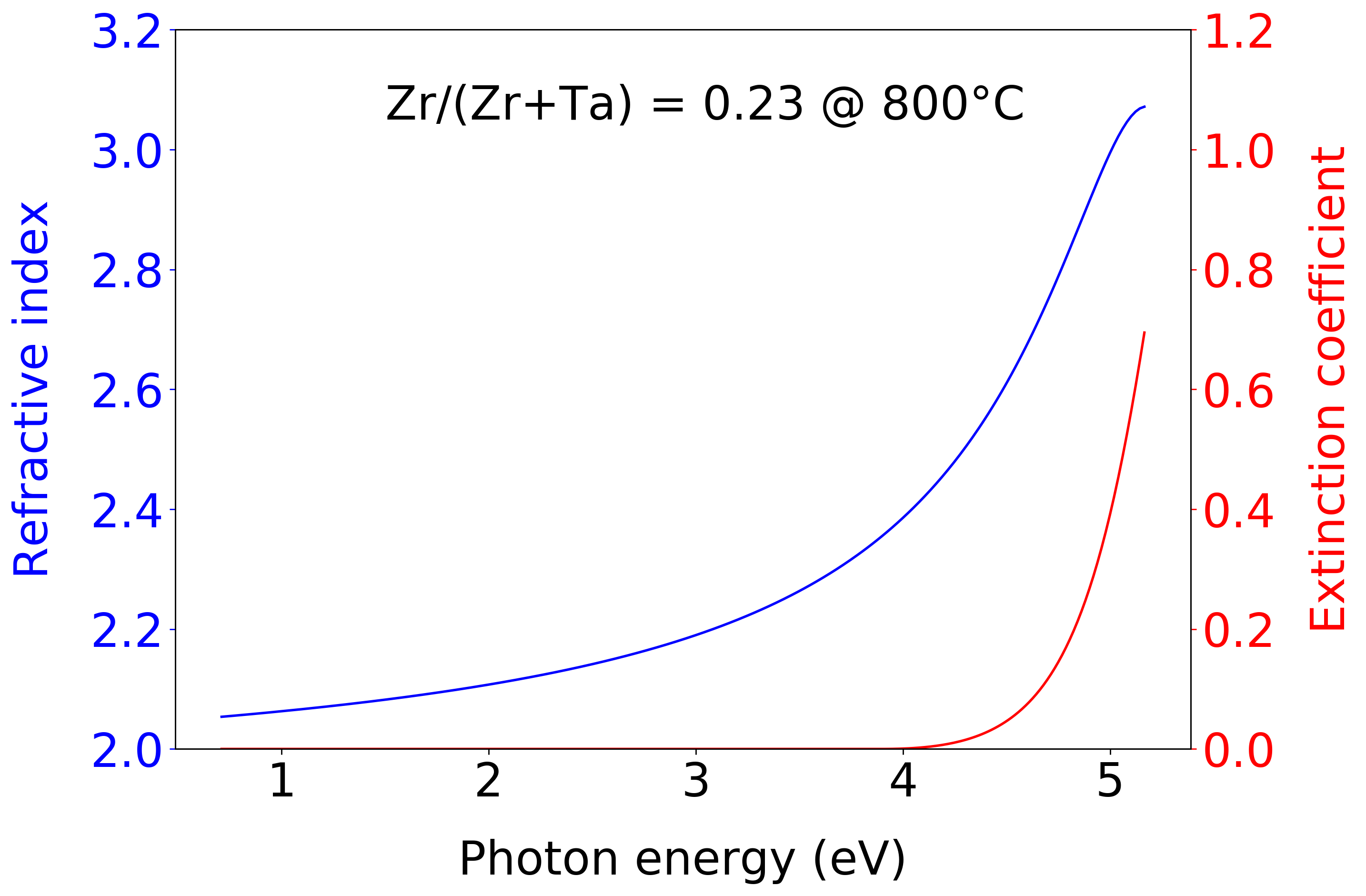}}
	\subfloat[][]{
		\includegraphics[width=0.45\linewidth]{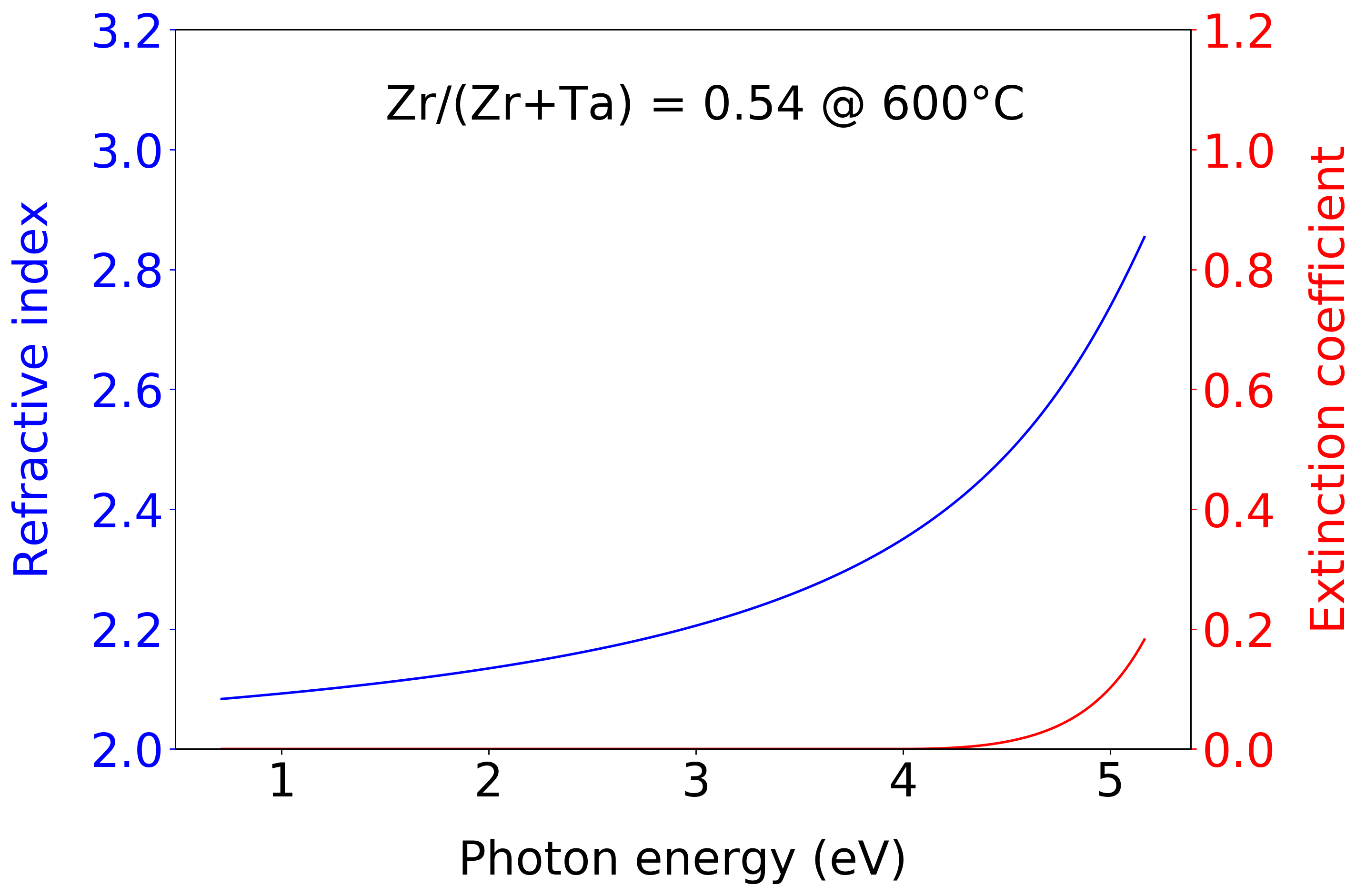}}
	\caption{\label{FIG_RBTD_n-k} Refractive index $n$ and extinction coefficient $k$ as function of photon energy for RBTD Ta$_2$O$_5$-ZrO$_2$ thin films: (a) cation ratio $\eta = 0.23$, annealed at 800 $^\circ$C for 10 hours; (b) cation ratio $\eta = 0.54$, annealed at 600 $^\circ$C for 10 hours.}
\end{figure}

The dispersion of refractive index $n$ and extinction coefficient $k$ are presented in Figure~\ref{FIG_RBTD_n-k}. The refractive index for both annealed films is presented in Table~\ref{TAB_optical_results}. The film with the largest cation ratio featured the highest refractive index, as expected, given that the refractive index of zirconia is higher than that of tantala.

Optical absorption at $\lambda$ = 1064 nm was determined by photothermal common-path interferometry \cite{alexandrovski2009photothermal}. We define the absorption normalized to a quarter wave thickness as
\begin{equation}
\label{EQ_alphaQW}
\alpha_{\textrm{\tiny{QWT}}} = \frac{\alpha}{d} \frac{\lambda}{4n}\ ,
\end{equation}
with $\alpha$ being the absorption as measured ppm, $n$ the refractive index, $d$ the coating thickness and $\lambda = 1064$ nm. $\alpha_{\textrm{\tiny{QWT}}}$ was 4.4 $\pm$ 0.8 for the film with cation ratio $\eta = 0.54$ and 1.85 $\pm$ 0.05 for the film with cation ratio $\eta = 0.23$. These results are summarized in Table \ref{TAB_optical_RBTD}.
\begin{table*}[!ht]
	\begin{center}
		\begin{tabular}{cccc}
			\hline
			Film & $T_a$ [$^{\circ}$] & Cation ratio $\eta$ & $\alpha_{\textrm{\tiny{QWT}}}$ @ 1064 nm ppm)\\
			\hline
			CSU II & 800 & 0.23 & 1.85 $\pm$ 0.05\\
			CSU III & 600 & 0.54 & 4.4 $\pm$ 0.8\\
			\hline
		\end{tabular}
	\end{center}
	\caption{Absorption loss normalized to a quarter wave thickness $\alpha_{\textrm{\tiny{QWT}}}$ at 1064 nm for RBTD Ta$_2$O$_5$-ZrO$_2$ thin films annealed at the maximum temperature $T_a$ before crystallization (see Sections \ref{SECT_annealing}. See Eq. (\ref{EQ_alphaQW}) for more details.}
	\label{TAB_optical_RBTD}
\end{table*}

\subsection{HiPIMS/RF-MS}
\label{SECT_MS_optical}
Similarly to what is described in Sections~\ref{SECT_LMAopt} and \ref{SECT_RBTDopt}, spectroscopic ellipsometry was carried out between wavelengths of 210 nm and 2500 nm with a J. A. Woollam Co., Inc. RC2-XI ellipsometer. Data analysis to obtain the complex refractive index and thickness mappings were carried out using the J. A. Woollam CompleteEASE software, with the optical properties being represented by a general oscillator model consisting of a Tauc-Lorentz and Gaussian oscillators for the UV absorption. The model parameters were first fit for each sample from a series of measurements at four different incidence angles, $45^{\circ}$, $55^{\circ}$, $65^{\circ}$ and $75^{\circ}$, carried out at the center of the sample. The average thickness of the samples was then established through a mapping of 48 points, measured at $70^{\circ}$ where only the thickness and surface roughness (Bruggeman effective medium approximation) were fit while keeping the optical model constant. The measured refractive index at 1064 and 1550 nm  of all HiPIMS/RF-MS samples is presented in Table~\ref{TAB_optical_results}.

\subsection{Comparative summary}
\label{SECT_opt_compar}
\begin{figure}
\centering
\includegraphics[width=0.45\linewidth]{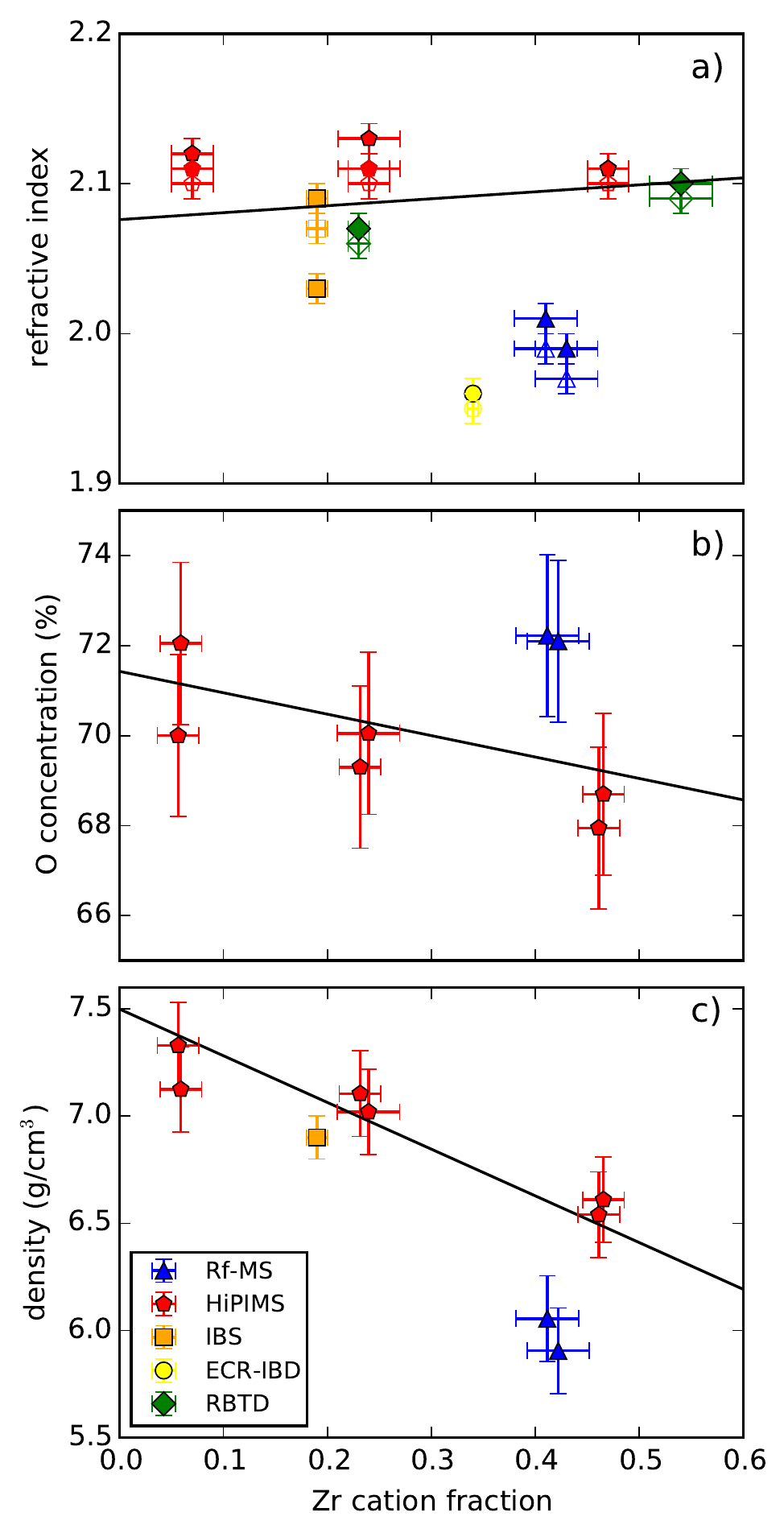}\hfill 
\caption{Correlation between oxygen concentration, density and refractive index of co-sputtered Ta$_2$O$_5$-ZrO$_2$ thin films, as a function of cation ratio $\eta =$ Zr/(Ta+Zr): (a) refractive index at 1064 nm (filled markers) and 1550 nm (empty markers), the black solid line is the value expected from the rule of mixtures at 1064 nm when considering the refractive index found in Ref. \cite{Wood:82} for bulk zirconia and Ref. \cite{bright} for that of bulk tantala; (b) oxygen concentration; (c) density. The solid lines in (b) and (c) are the expected oxygen content and density, respectively,  according to the rule of mixture and assuming stoichiometric tantala and zirconia for the oxygen content, and density values for Ta$_2$O$_5$ and ZrO$_2$ taken from the literature \cite{Vajente_2018,ZrO2density}. Red pentagons and blue triangles denote HiPIMS and RF-MS samples, respectively, while yellow circle, orange squares and green diamonds denote ECR-IBD, IBS (LMA) and RBTD samples, respectively.
}\label{FIG_compar_density_O_vs_conc}
\end{figure}

Figure~\ref{FIG_compar_density_O_vs_conc}a shows the refractive index values at 1064 nm and 1550 nm reported in Table~\ref{TAB_optical_results} as a function of the cation ratio $\eta =$ Zr/(Zr+Ta). The solid line corresponds to the rule of mixtures, assuming bulk values for pure oxides. The low dependence on $\eta$ is expected, since tantala and zirconia have a similar refractive index at these wavelengths. Samples deposited by HiPIMS, IBS and RBTD feature values close to the rule of mixture. For samples where tantala is deposited by RF-MS or when the coating is deposited by ECR-IBD, the refractive index is significantly lower.

In Figure~\ref{FIG_compar_density_O_vs_conc}b, we find that the oxygen content in the layers where tantala is deposited by HiPIMS follows the rule of mixture between stoichiometric tantala and zirconia, while there is an excess of oxygen in the films where tantala was deposited by RF-MS. Accordingly, as shown by Figure~\ref{FIG_compar_density_O_vs_conc}c, the density of films where tantala is deposited by HiPIMS or IBS follow the rule of mixture considering bulk densities of tantala and zirconia \cite{Vajente_2018,ZrO2density}, while those where tantala was deposited by RF-MS show a significantly lower density. Hence, the excess of oxygen leads to films with lower density and correspondingly lower refractive index. The explanation partly lies in the fact that atoms, when deposited by HiPIMS, have a higher kinetic energy than in RF-MS, which produces more compact films with less chances to have, for example, two oxygen atoms on the same bond. It is worth noting that we did not attempt to grow stoichiometric RF-MS layers, for example by changing the O$_2$:Ar ratio during deposition, as we wanted to compare HiPIMS and RF-MS under the same conditions.

Our results show the correlation between the oxygen concentration, the density, and the refractive index. Techniques such as HiPIMS and IBS feature atomic fluxes with higher kinetic energy, leading to higher dissociation rates of oxygen. This means that less oxygen may be used in the Ar/O2 mixture to obtain stoichiometry. Also, the higher energy of species impinging on the growing film surface leads to coatings of higher density. Conversely, RF-MS films are generally less dense and have a lower refractive index. However, pores are frequently filled with water vapour that adds oxygen and hydrogen, leading to apparent oxygen over-stoichiometry. The former may therefore have more desirable properties for optical components. We will see in Section~\ref{SECT_mech} how that affects the loss angle.
%
\section{Mechanical properties}
\label{SECT_mech}
The mechanical loss $\varphi_c$, Young modulus $Y_c$ and Poisson ratio $\nu_c$ of the coating materials were measured using different setups, all based on the ring-down technique \cite{Nowick72}: the mechanical resonances (modes) of a suspended coated resonator are excited using an actuator, and the amplitude of their ring-down oscillations is monitored using an optical transducer. For a resonating mode of frequency $f$ and ring-down time $\tau$, the measured mechanical loss angle is $\varphi = (\pi f \tau)^{-1}$. The coating loss angle $\varphi_c$ can be calculated as
\begin{equation}
\label{EQ_coatLoss}
\varphi_c = \frac{\varphi + (D-1)\varphi_s}{D} \ ,
\end{equation}
where $\varphi_s$ is the measured loss angle of the substrate, $\varphi$ is the measured loss angle of the coated sample and $D$ is a so-called \textit{dilution factor}, defined as the ratio of the coating elastic energy to the total elastic energy of the coated sample.

Coated samples were suspended under vacuum, in order to prevent systematic damping from ambient pressure. We used different suspension systems, each one specifically adapted to the different geometries of the substrates used as resonators.  Because of the different suspension systems and samples, we used also different methods to estimate the dilution factor and the coating Young modulus and Poisson ratio. Specific measurements and their analyses are described in the following.

Results are summarized in Table \ref{TAB_mech_results}, where an average coating loss $\overline{\varphi}$, taken from every set of resonant modes of each sample, is used to ease comparisons. A comparative summary is presented in Figures \ref{FIG_averaged-loss-angle-doping} and \ref{FIG_averaged-loss-angle-annealing} of Section \ref{SECT_mech_compar}, where the effects of different cation ratios $\eta =$ Zr/(Zr+Ta) and annealing temperatures $T_a$ on the coating mechanical loss are discussed.  
\begin{table*}[!ht]
	\begin{center}
		\begin{tabular}{ccccccc}
			\hline
			Sample & Process & $T_a$ [$^\circ$C] & $\eta$ [at.\%] & $\overline{\varphi}$ [$10^{-4}$ rad] & $Y_c$ [GPa] & $\nu_c$ \\
			\hline
            MLD2014 & IBS & 800 & $48.5 \pm 0.4$ & $4.26 \pm 0.11$ & $130.5\pm 2$\\
            MLD2018 & IBS & 700 & $50.2 \pm 0.6$ & $1.82 \pm 0.04$\\
            LMA & IBS & 700 & 19 $\pm$ 1 & 4.41 $\pm$ 0.68 & 120 $\pm$ 3 & 0.32 $\pm$ 0.01\\
            UoS & ECR IBD & 750 & $34.0 \pm 0.5$ & 3.64 $\pm$ 0.45 & $176 \pm 8$\\
            UMP 551 & RF-MS & 800 & 41 $\pm$ 3 & 2.36 $\pm$ 0.85 & 125 $\pm$ 2 & 0.21 $\pm$ 0.05\\ 
            UMP 554 & RF-MS & 800 & 43 $\pm$ 3 & 2.52 $\pm$ 0.82 & 114 $\pm$ 2 & 0.36 $\pm$ 0.03\\ 
            UMP 658 & HiPIMS & 700 & 24 $\pm$ 3 & 5.59 $\pm$ 0.90 & 110.6 $\pm$ 1.2 &  0.31 $\pm$ 0.02\\
            UMP 659 & HiPIMS & 700 & 24 $\pm$ 2 & 5.04 $\pm$ 1.53 & 108 $\pm$ 5 & 0.35 $\pm$ 0.05\\ 
            UMP 542 & HiPIMS & 600 & 7 $\pm$ 2 & 5.69 $\pm$ 0.37 & 100.6 $\pm$ 1.1 & 0.34 $\pm$ 0.02\\ 
            UMP 664 & HiPIMS & 650 & 7 $\pm$ 2 & 4.45 $\pm$ 0.38 & 116.5 $\pm$ 1.3 & 0.33 $\pm$ 0.02\\ 
            UMP 678 & HiPIMS & 800 & 47 $\pm$ 2 & 4.38 $\pm$ 0.63 & 111 $\pm$ 4 & 0.28 $\pm$ 0.09\\ 
            UMP 680 & HiPIMS & 800 & 47 $\pm$ 2 & 4.87 $\pm$ 0.92 & 110 $\pm$ 4 & 0.28 $\pm$ 0.09\\
            CSU II & RBTD & 800 & 23 $\pm$ 1 & 3.22 $\pm$ 0.53 & 145 $\pm$ 6 & 0.30 $\pm$ 0.09\\
            CSU III & RBTD & 600 & 54 $\pm$ 3 & 2.67 $\pm$ 0.15 & 143 $\pm$ 5 & 0.37 $\pm$ 0.05\\
			\hline
		\end{tabular}
	\end{center}
	\caption{Measured mechanical properties of co-sputtered Ta$_2$O$_5$-ZrO$_2$ thin films: frequency-averaged mechanical loss $\overline{\varphi}$ of samples annealed at the highest peak temperature $T_a$ before the onset of crystallization, Young modulus $Y$, Poisson ratio $\nu$. RF-MS samples had Ta deposited by RF-MS and Zr deposited by DC-MS, HiPIMS samples had Ta deposited by HiPIMS and Zr deposited by RF-MS. RF-MS, HiPIMS and IBS (LMA) samples have been annealed for 10 hours, ECR-IBD samples for 5 hours. Average coating loss values are plotted in Figures \ref{FIG_averaged-loss-angle-doping} and \ref{FIG_averaged-loss-angle-annealing} and discussed in Section \ref{SECT_mech_compar}.}
	\label{TAB_mech_results}
\end{table*}

\subsection{IBS}
\subsubsection{MLD --}\label{SEC_MLDmech}
The coating samples used to measure coating mechanical loss, Young modulus and Poisson ratio were grown on  fused-silica disks and cantilevers and on silicon cantilevers (see Section \ref{SECT_MLDsamples}). Samples grown on fused-silica substrates were used for measurements at ambient temperature, those on silicon cantilevers to perform cryogenic measurements.

{\it Disks --} The mechanical loss measurements were performed at Hobart and William Smith Colleges, where the disks were welded to monolithic fused-silica fibers and suspended. The mechanical loss of the lowest eight vibrational modes of the disks, which span the frequencies from 2.5 to 18 kHz, were measured by resonating the samples using an electrostatic comb exciter and then measuring the ring-down using an ellipsometry readout \cite{Harry2006}. Each mode was measured several times with both weighted and unweighted averages retained.

After each set of measurements, the samples have been annealed and remeasured. The annealing temperatures were 300, 600, 650, 700, 750 and 800 $^\circ$C for MLD2014 and 400, 600 and 700 $^\circ$C for MLD2018. For these experiments, the fused silica substrate is typically annealed to high temperature before being coated. The MLD2018 disk was properly annealed at 950 $^\circ$C, whereas the MLD2014 disk was not annealed prior to coating. For the latter sample, the substrate loss was determined from the data using the frequency dependence of the loss as determined by finite-element modeling. The model, created using COMSOL, was used to calculate the dilution factor for the total coating loss and for the decomposition of the coating loss into bulk and shear loss \cite{Hong:2012}.

\begin{figure}[h!]
	\centering
	\subfloat[]{\includegraphics[width=0.45\textwidth]{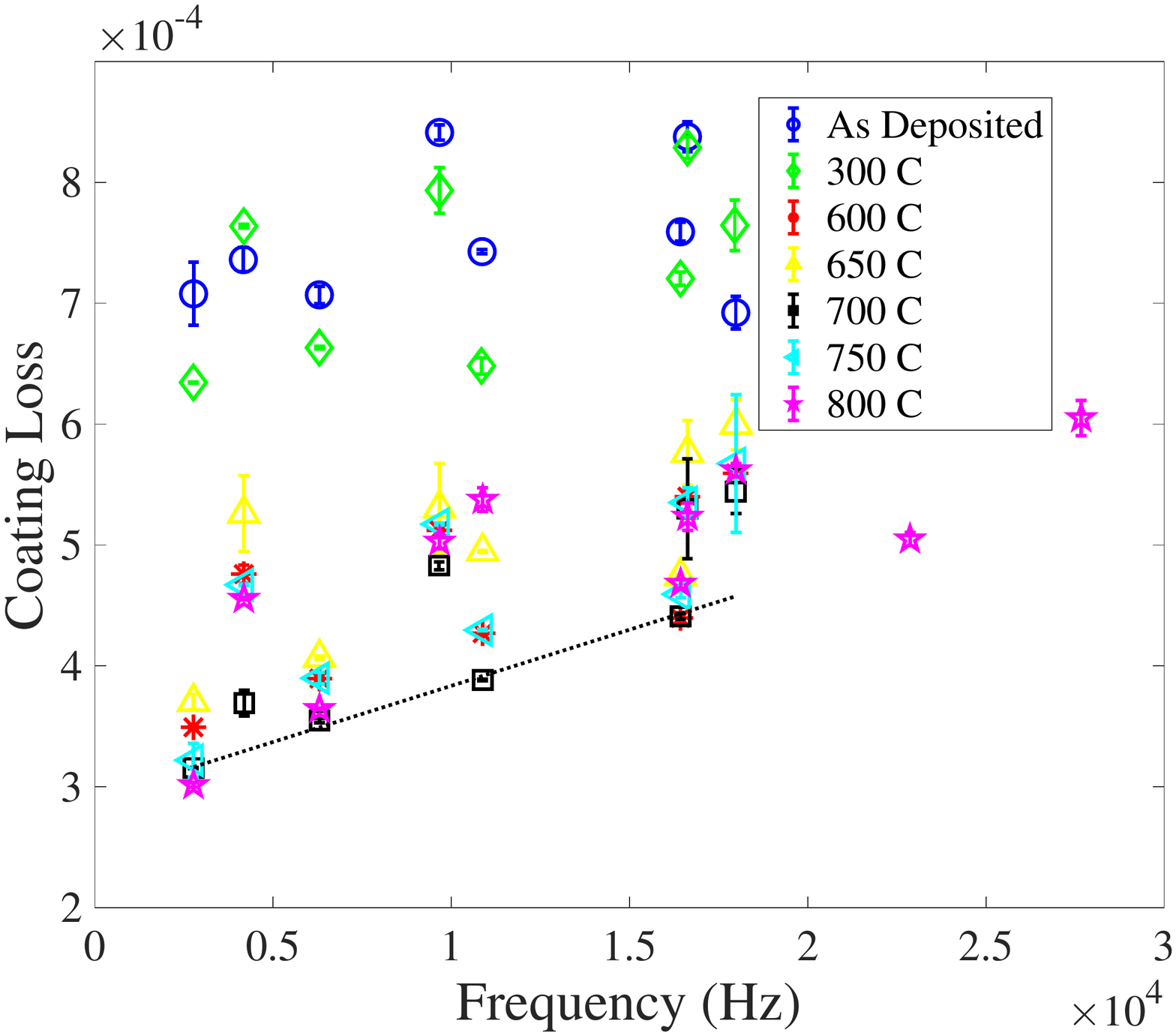}}
	\subfloat[]{\includegraphics[width=0.45\textwidth]{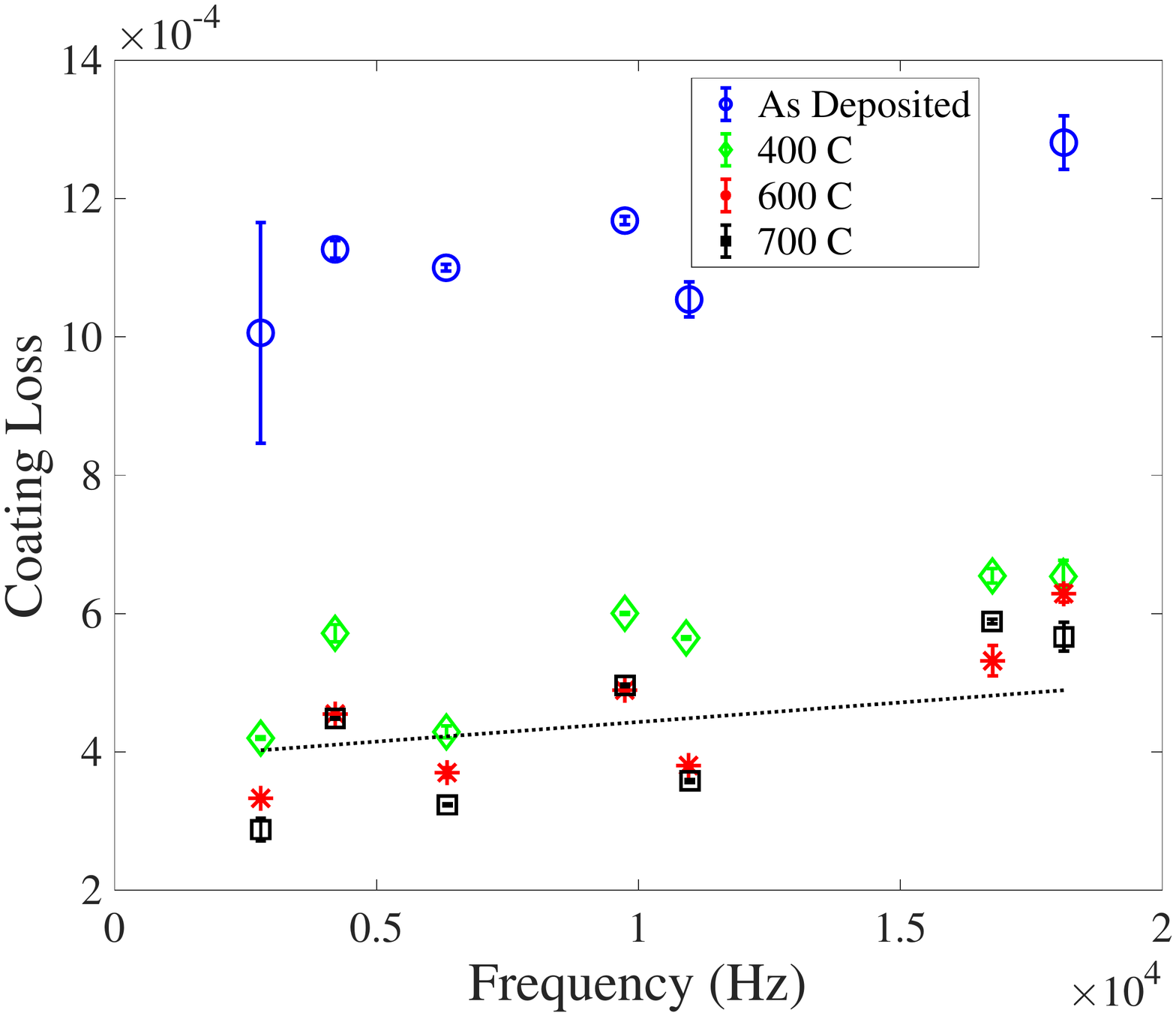}}
	\caption{Coating loss as a function of frequency for IBS Ta$_2$O$_5$-ZrO$_2$ thin films, as measured on suspended disks: (a) sample MLD2014, (b) sample MLD2018. The loss is steadily reduced with increased annealing temperature $T_a$, until reaching a minimum near 700 $^\circ$C. A linear fit is shown (black line) for the data sets of the 700-$^\circ$C annealing.}
	\label{FIG_MLDcoatingLoss}
\end{figure}

As shown in Figure~\ref{FIG_MLDcoatingLoss}, the coating loss for both samples decreased with increasing annealing temperature, until reaching a minimum at 700 $^\circ$C. A linear fit of the frequency-dependent coating loss after annealing at 700 $^\circ$C gave:

$\varphi_c\left( f \right) = \left(9.29 \pm 0.09 \right)\times 10^{-9}\, \mathrm{Hz}^{-1} \,\,f + \left(2.907 \pm 0.007\right)\times 10^{-4} $ for MLD2014 

$\varphi_c\left( f \right) = \left(5.7 \pm 0.2 \right)\times 10^{-9}\,\mathrm{Hz}^{-1}\,\,f +  \left( 3.87  \pm 0.02  \right)\times 10^{-4} $ for MLD2018.

\begin{figure}[h!]
	\centering
	\subfloat[]{\includegraphics[width=0.45\textwidth]{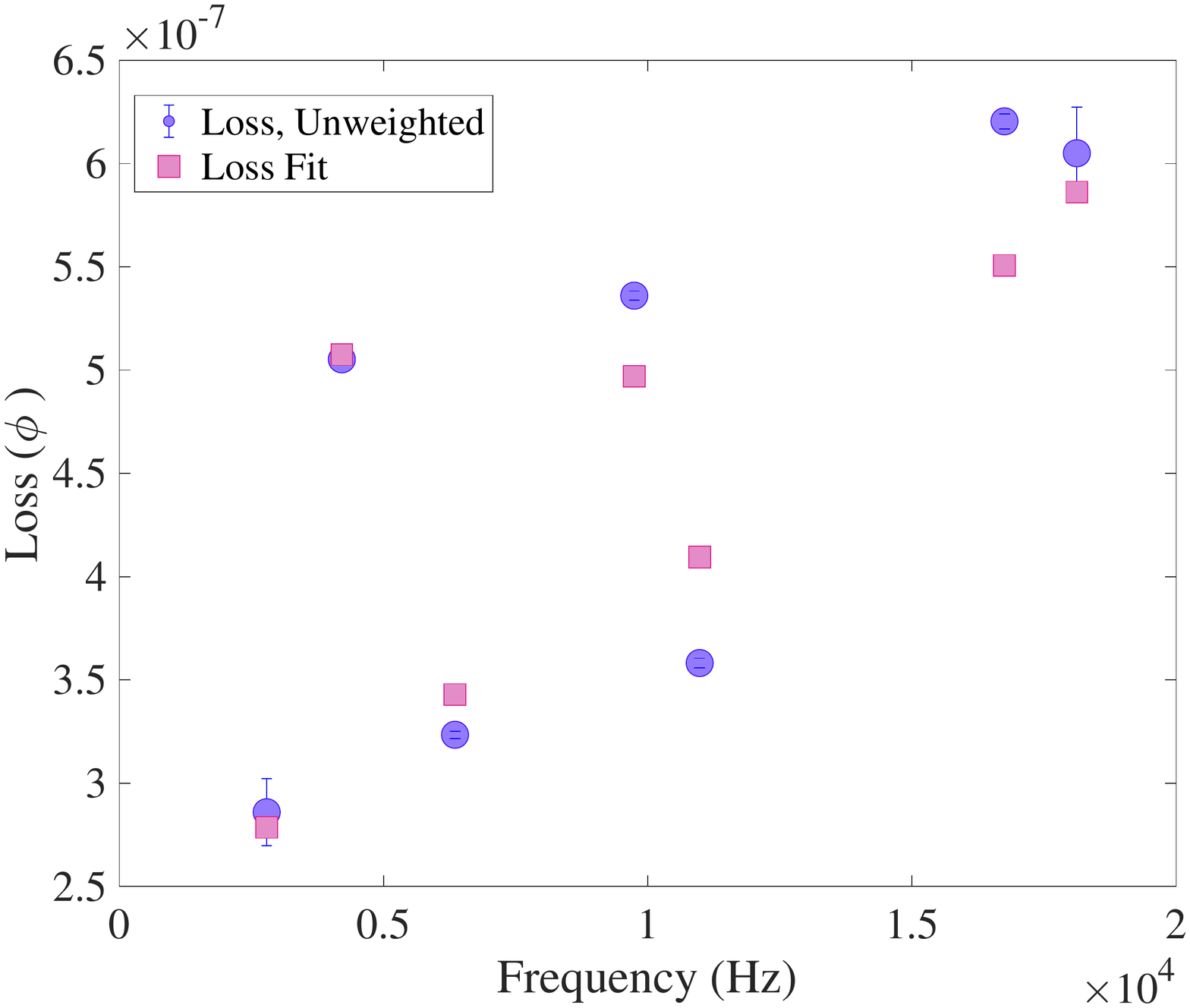}}
	\subfloat[]{\includegraphics[width=0.45\textwidth]{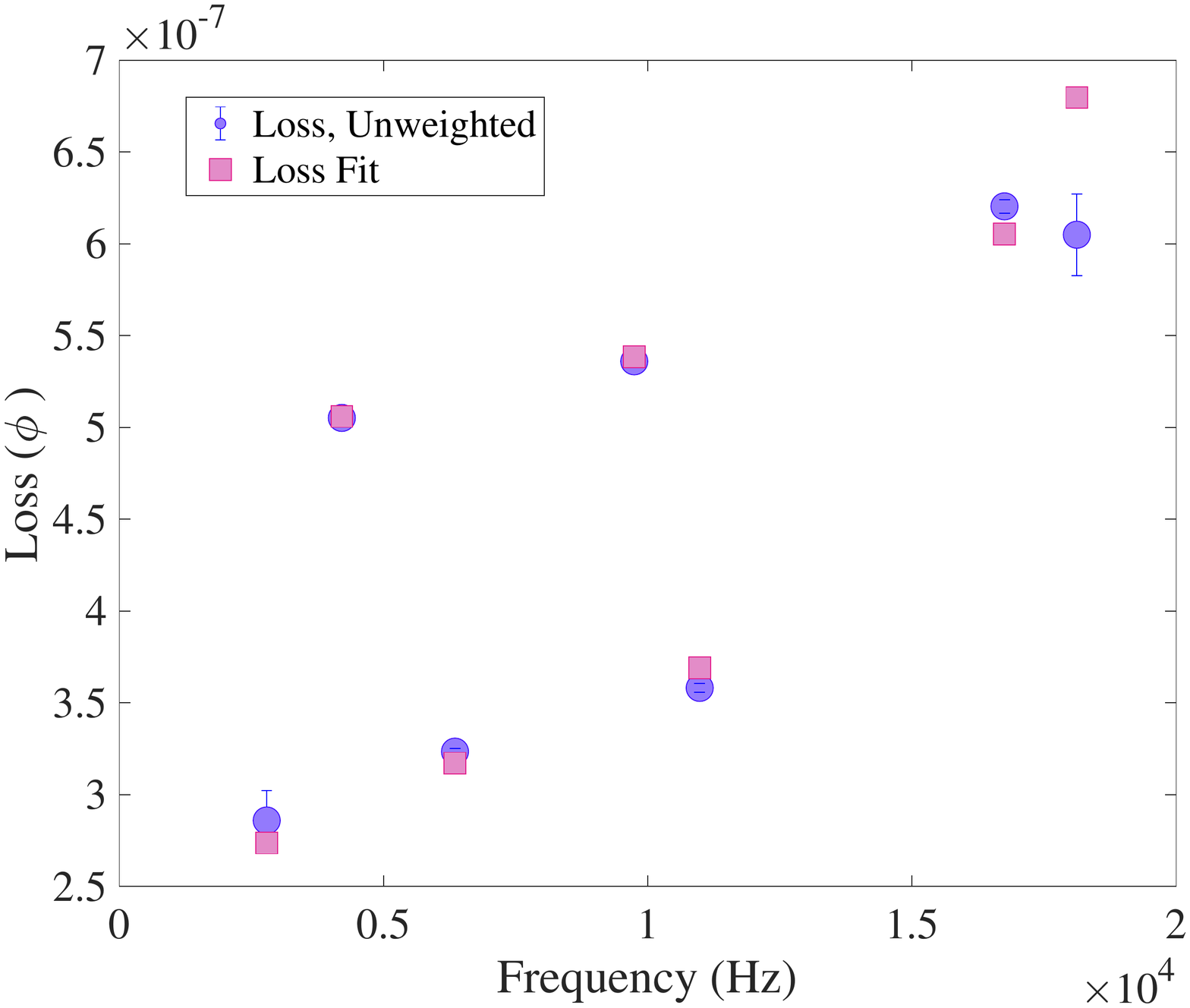}}
	\caption{Comparison of analysis methods, as applied to coating loss data of IBS Ta$_2$O$_5$-ZrO$_2$ thin films (MLD) measured on suspended disks: (a) fit with single loss angle, (b) fit with bulk and shear loss angles. The need for separate analysis of bulk and shear coating loss angle is pointed out by the improved fit of panel (b).}
	\label{FIG_MLD_bulkShearExample}
\end{figure}

\begin{figure}[h!]
	\centering
	\subfloat[]{\includegraphics[width=0.45\textwidth]{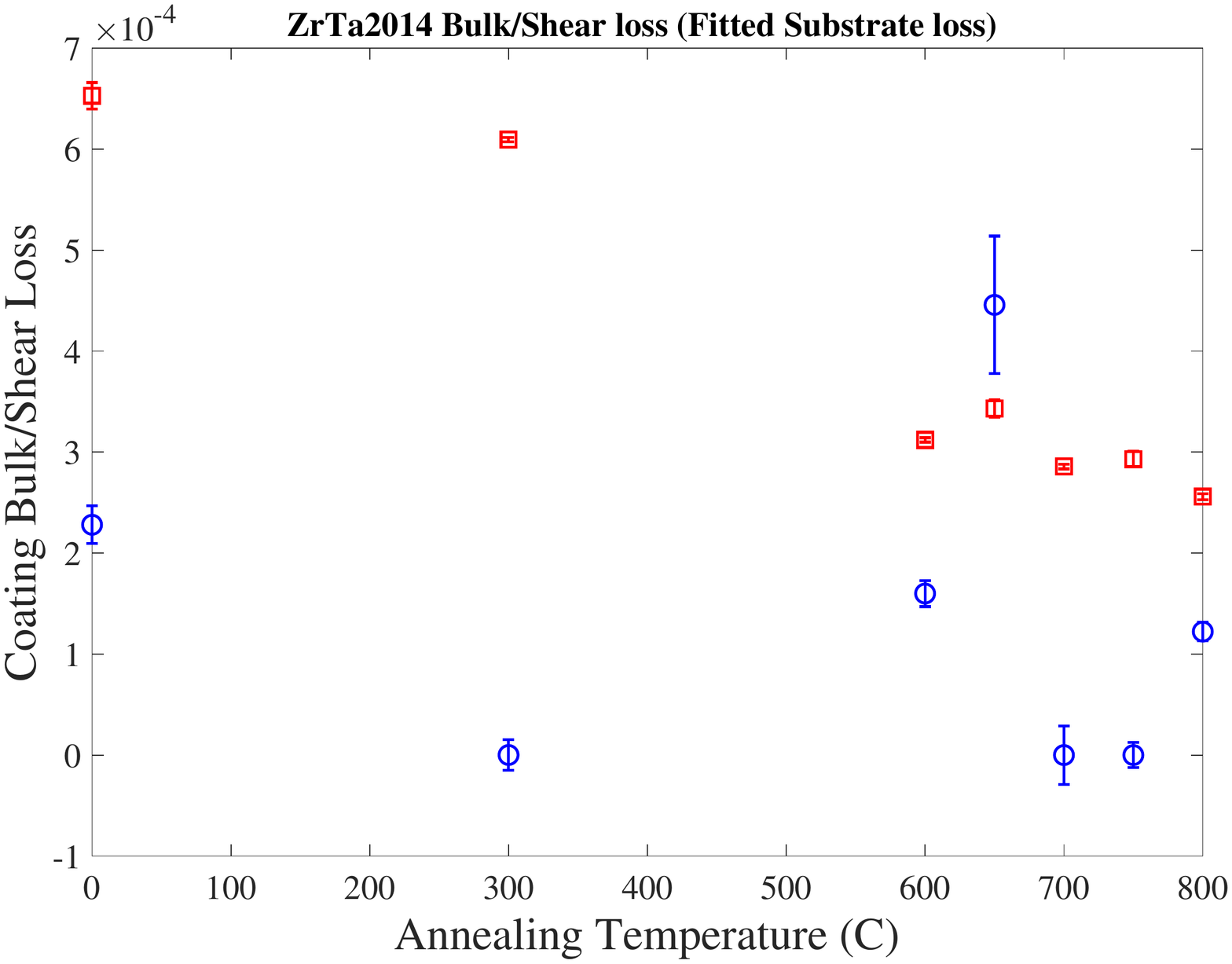}}
	\subfloat[]{\includegraphics[width=0.45\textwidth]{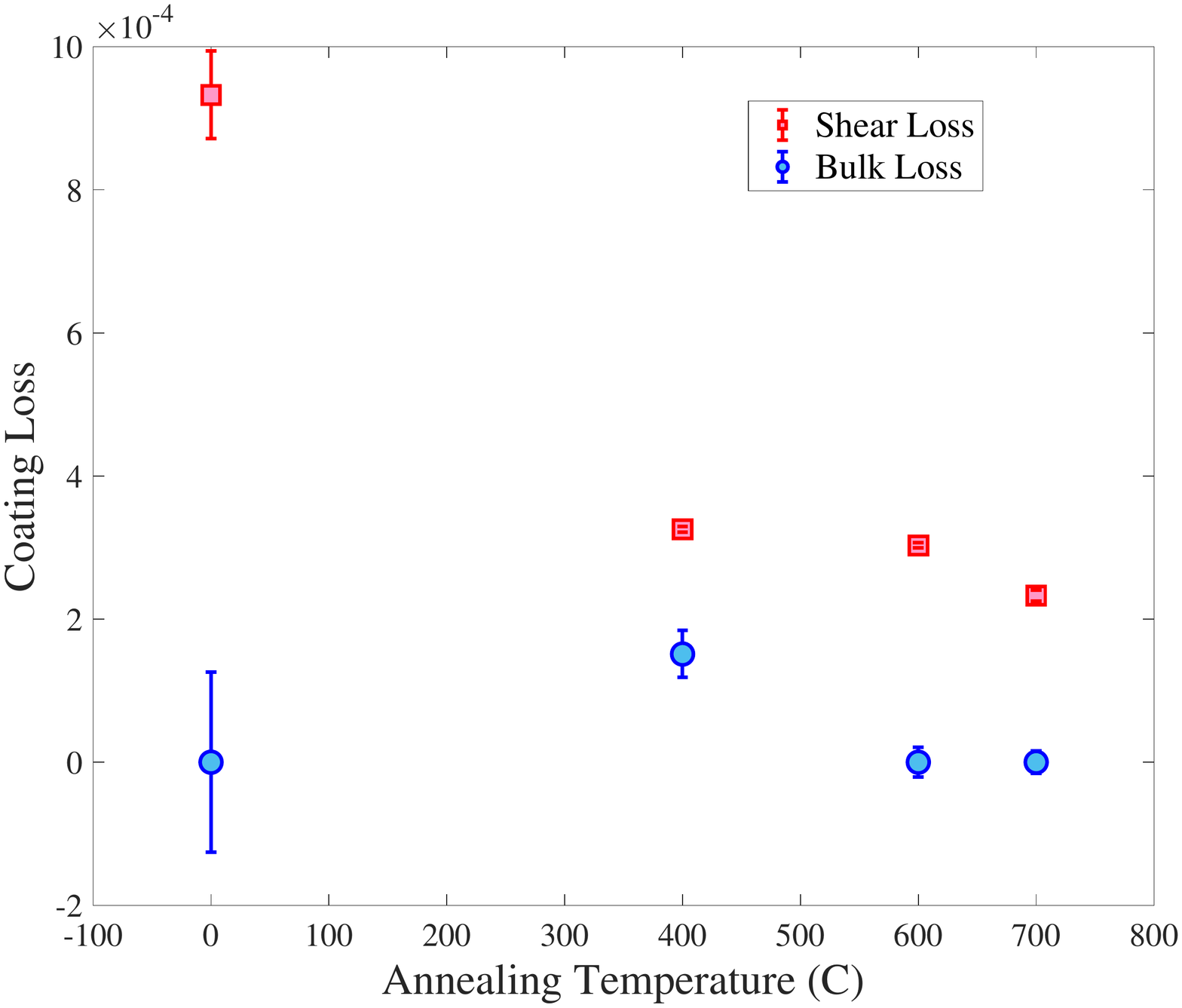}}
	\caption{Bulk and shear coating loss as function of annealing temperature for IBS Ta$_2$O$_5$-ZrO$_2$ thin films, as measured on suspended disks: (a) sample MLD2014, (b) sample MLD2018. The shear loss is steadily reduced with increased annealing temperature, while the bulk loss shows no clear dependence.}
	\label{FIG_MLD_bulkShearLoss}
\end{figure}

The variation in the ratio of bulk and shear energy with mode shape, which in turn depends on frequency, introduces a scatter of the data about any function of a single coating loss angle. Instead, a decomposition of the coating loss into shear and bulk components allows a more precise fit to the data, as shown in Figure~\ref{FIG_MLD_bulkShearExample}. A fit of each data set to bulk and shear coating loss resulted in the dependence of these values with annealing temperature $T_a$ shown in Figure~\ref{FIG_MLD_bulkShearLoss}: the shear loss steadily decreased for increasing $T_a$, whereas the bulk loss showed no clear dependence.

The coating Young modulus was measured after annealing at two different temperatures, using a nano-indentation facility at the California Institute of Technology. The results, $Y = 130 \pm 2$ GPa for MLD2018 annealed at 800 $^\circ$C and $Y = 131 \pm 2$ GPa for MLD2014 annealed at 600 $^\circ$C, assume a coating Poisson ratio of 0.298 determined by molecular dynamics models \cite{Trinastic}.

{\it Cantilevers --}
The mechanical loss of the MLD2018 coating was measured also on a fused-silica cantilever at ambient temperature, and on silicon cantilevers at cryogenic temperatures. Measurements were performed at the Institute for Gravitational Research of the University of Glasgow.

The cantilevers were mounted in a clamp, under vacuum, and their bending modes were excited using an electrostatic actuator. The amplitude of the ring-down motion was monitored using an optical shadow sensor. The coating mechanical loss was then calculated using Eq. (\ref{EQ_coatLoss}) and
\begin{equation}
\label{EQ_dilFact_cantilev}
D = \frac{3 Yd}{Y_s d_s}\ ,
\end{equation}
where $Y$ and $d$ are the coating Young modulus and thickness, respectively, and $Y_s$ and $d_s$ are the substrate Young modulus and thickness, respectively.

\begin{figure} 
\centering
	\includegraphics[width=0.75\textwidth]{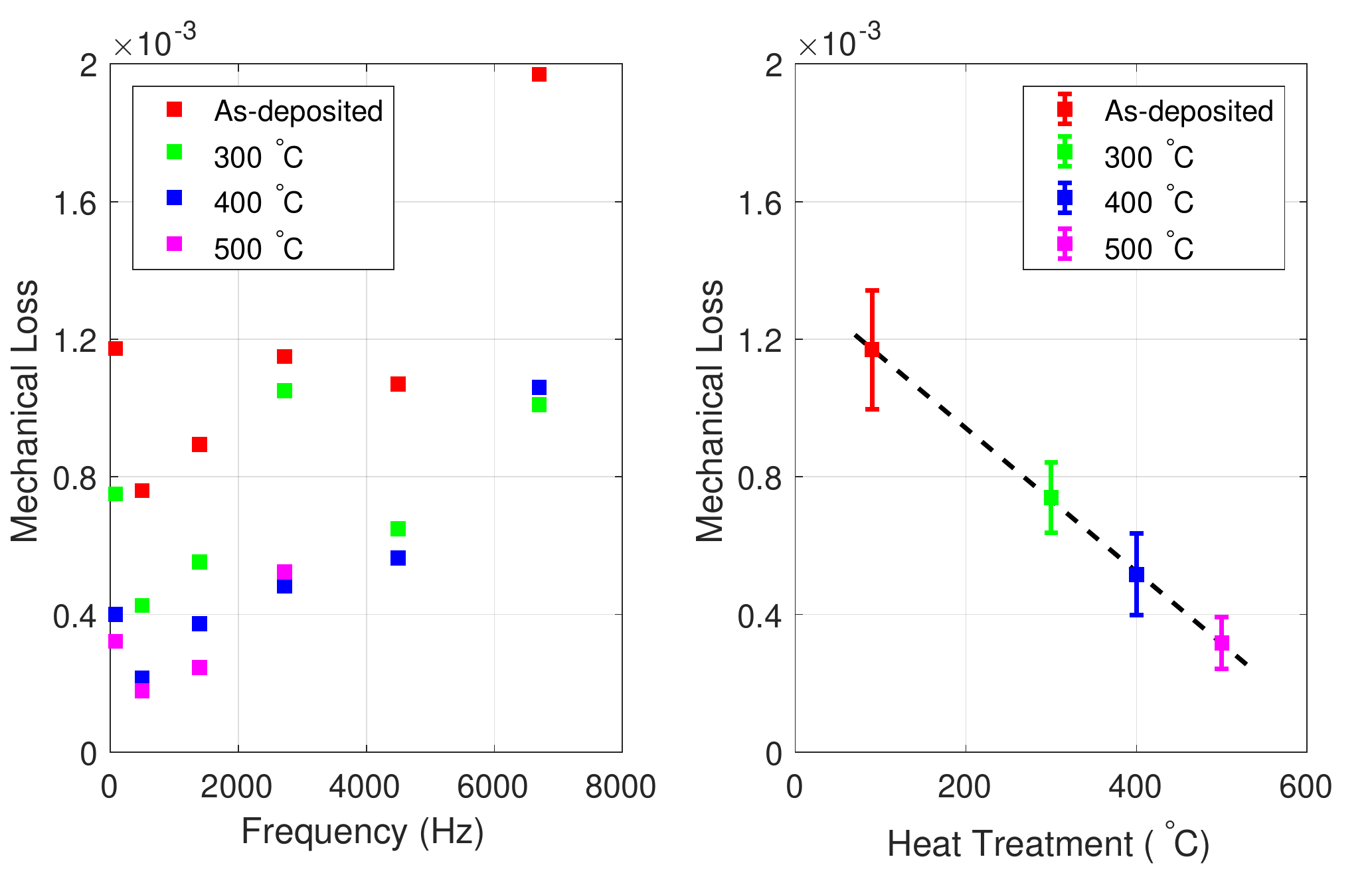}
	\caption{Coating loss of IBS Ta$_2$O$_5$-ZrO$_2$ thin film MLD2018, as measured on a clamped $\sim$100-$\mu$m thick fused-silica cantilever at ambient temperature: (a) the loss was measured as a function of frequency on the as-deposited sample, then re-measured after progressive annealings at 300, 400 and 500 $^{\circ}$C; (b) frequency-averaged coating loss $\overline{\varphi}$ as a function of annealing temperature. The coating loss decreased after each annealing treatment.}
	\label{FIG_MLD_cantilev_RT}
\end{figure}

After each set of measurements, the coated fused-silica cantilever was progressively annealed at increasing peak temperatures, up to 500 $^{\circ}$C. Figure \ref{FIG_MLD_cantilev_RT} shows the effect of annealing on the coating mechanical loss as a function of frequency, and that the coating mechanical loss at $1$~kHz decreased by over a factor of three upon annealing, from $\overline{\varphi} = 1.1\times10^{-3}$ in its as-deposited state to $\overline{\varphi} = 3.2\times10^{-4}$ after a 500$^{\circ}$C annealing.

\begin{figure}[h] 
\centering
	\includegraphics[width=0.75\textwidth]{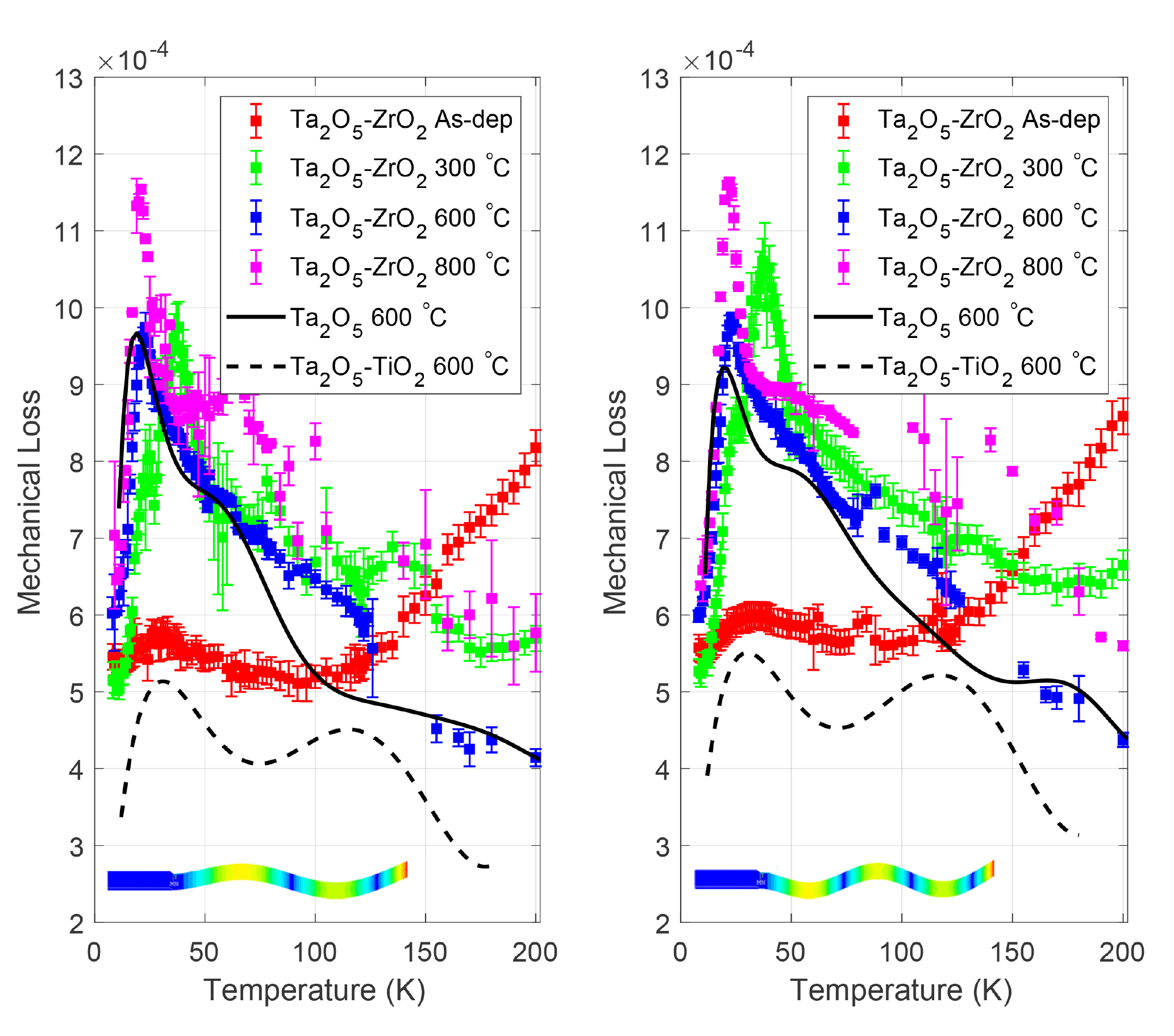}
	\caption{Cryogenic mechanical loss of IBS Ta$_2$O$_5$-ZrO$_2$ coatings as a function of sample temperature, as measured on clamped $\sim$65-$\mu$m thick silicon cantilevers: (a) results for the third-order bending mode at $\sim$1.4 kHz, (b) results for the fourth-order bending mode at $\sim$2.7 kHz. The coating loss of sample MLD2018 as deposited and after a 300-$^{\circ}$C annealing and of sample MLD2014 after 600-$^{\circ}$C and 800-$^{\circ}$C annealing are shown, and compared to data of IBS amorphous Ta$_2$O$_5$ and poly-crystalline Ta$_2$O$_5$-TiO$_2$ samples annealed to 600 $^{\circ}$C \cite{Martin_2010,Titania_2021}. Finite-element representations of the measured resonant mode are visible at the bottom of each plot.}
	\label{FIG_MLD_cantilev_cryoT}
\end{figure}

The coated silicon cantilevers were measured in a temperature-controlled, liquid-helium-cooled cryostat \cite{RobieThesis}. The coating mechanical loss was calculated using Eqs. (\ref{EQ_coatLoss}) and (\ref{EQ_dilFact_cantilev}) for a large number of resonant modes, from $\sim$1 to 30 kHz, as extensively detailed in \cite{RobieThesis}. Here, for sake of conciseness, we present in Figure \ref{FIG_MLD_cantilev_cryoT} only the coating loss of the third- and fourth-order bending modes, at $\sim$1.4 kHz and $\sim$2.7 kHz respectively.

Figure \ref{FIG_MLD_cantilev_cryoT} shows that IBS Ta$_2$O$_5$-ZrO$_2$ coating in its as-deposited state has a very small and broad loss peak of $\sim$6$\times10^{-4}$ at $\sim$30 K. This peak narrowed to $1\times10^{-3}$ after annealing at 300 and 400 $^{\circ}$C, and further increased to $1.2\times10^{-3}$ after an 800 $^{\circ}$C anneal. Below 100 K, the coating loss temperature-dependent trend of the annealed Ta$_2$O$_5$-ZrO$_2$ samples is similar to that of pure IBS Ta$_2$O$_5$ films annealed at 600 $^{\circ}$C \cite{Martin_2010}. After annealing at 800 $^{\circ}$C, the coating loss of Ta$_2$O$_5$-ZrO$_2$ does not show the large and broad loss peak observed at 80 K for IBS poly-crystalline Ta$_2$O$_5$ thin films annealed to the same temperature \cite{Martin_2010}. This is a clear indication that co-sputtering ZrO$_2$ with Ta$_2$O$_5$ prevented the onset of crystallization for annealing at 800 $^{\circ}$C.

As shown in Figure \ref{FIG_MLD_cantilev_RT}, annealing reduced the coating  loss at ambient temperature. At the same time, Figure \ref{FIG_MLD_cantilev_cryoT} suggests that annealing increased the coating loss at low temperatures, which is the opposite of what has been observed on IBS poly-crystalline Ta$_2$O$_5$-TiO$_2$ \cite{Titania_2021}. This is of interest for further understanding of coating loss and structure and modelling, and suggests that different values of the cation ratio $\eta=$ Zr/(Zr+Ta) need to be investigated at low temperatures.

\subsubsection{LMA --}
\begin{figure}
\centering
	\subfloat[]
	{\includegraphics[height=0.35\textwidth]{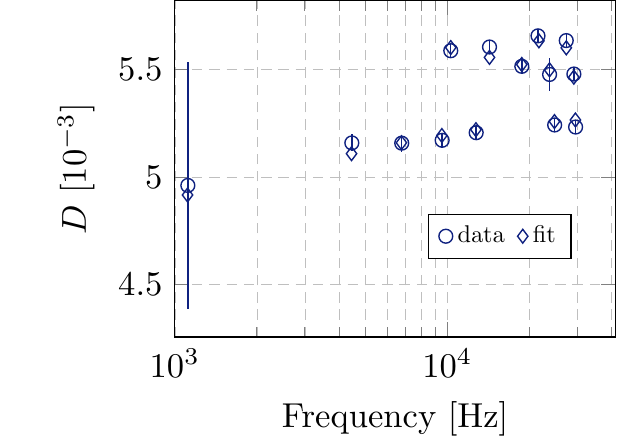}\label{FIG_LMA_D}}\quad
	\subfloat[]
	{\includegraphics[height=0.35\textwidth]{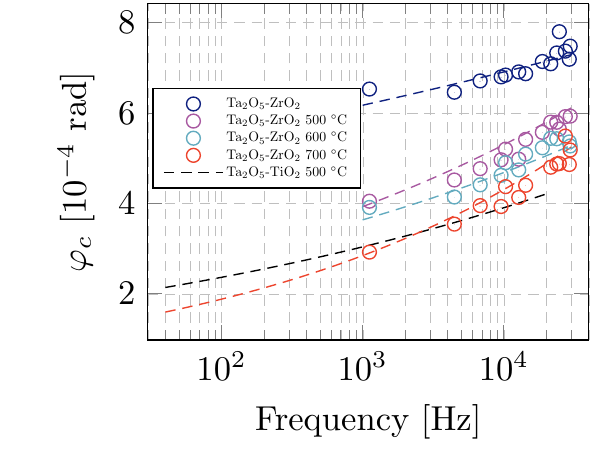}\label{FIG_LMA_coatLoss}}
	\caption{Measured mechanical properties of IBS Ta$_2$O$_5$-ZrO$_2$ thin films (LMA) as a function of frequency, as measured on a disk suspended with a GeNS system: (a) measured and best-fit simulated dilution factor $D$, for the first three families of disk mode shapes \cite{Granata20}; (b) coating loss angle $\varphi_c$, as measured after deposition and after annealing treatments at 500, 600 and 700 $^\circ$C for 10 hours, in air. Data of the annealed Ta$_2$O$_5$-TiO$_2$ layers of Advanced LIGO and Advanced Virgo is also shown, for comparison \cite{Granata20}. Dashed lines are least-squares fits using a power-law model $\varphi_c(f) = af^b$, best-fit parameters $a$ and $b$ are presented in Table \ref{TAB_LMAfit}. Values at $f<1$ kHz are model extrapolations.}
\end{figure}

We used a Gentle Nodal Suspension (GeNS) system \cite{Cesarini09} to measure coating mechanical loss, Young modulus and Poisson ratio via the direct measurement of the disk dilution factor \cite{Granata20}. As a matter of fact, the dilution factor $D$ in Eq. (\ref{EQ_coatLoss}) can be written as a function of the resonator mode frequency ($f_0$, $f$) and mass ($m_0$, $m$) before and after coating deposition, respectively \cite{Li14}:
\begin{equation}
\label{EQ_dilFact_disks}
D = 1 -  \left( \frac{f_0}{f} \right)^2 \ \frac{m_0}{m} \ .
\end{equation}
The advantage of measuring the dilution factor is that the ensuing coating loss estimation is then exclusively based on measured quantities: sample loss angle, mode frequency and mass. Therefore, unlike other experimental setups based on the ring-down method, our method does not require prior knowledge of coating Young modulus and thickness. Furthermore, one can estimate coating Young modulus $Y$ and Poisson ratio $\nu$ by fitting the results of finite-element simulations to measured values of $D$ estimated via Eq. (\ref{EQ_dilFact_disks}). Compared to nano-indentation, such method does not rely on the prior knowledge of $\nu$, and does not depend on the nature of the substrate used nor on the model chosen for data analysis \cite{Granata20}. Figure~\ref{FIG_LMA_D} shows best-fit results for three mode families of different shapes and frequency \cite{Granata20}, giving $Y = 120 \pm 3$ GPa and $\nu = 0.32 \pm 0.01$. Details about our fitting method and finite-element simulations can be found elsewhere \cite{Granata20,Granata16}.

Figure \ref{FIG_LMA_coatLoss} shows the coating loss angle before and after consecutive annealing treatments at 500, 600 and 700 $^\circ$C for 10 hours, in air. As expected, the annealing reduced the coating loss. However, the coating showed cracks after the 700 $^\circ$C annealing. This issue is further discussed in Section \ref{SECT_annealing}.

We used a power-law model $\varphi_c(f) = af^b$ to describe the observed frequency-dependent behavior of the coating internal friction \cite{Gilroy81,Travasso07,Cagnoli18}, and applied least-squares linear regression in order to fit the data. Best-fit parameters are presented in Table \ref{TAB_LMAfit}. It is worth noting that the $b$ parameter increased after each annealing step, implying lower loss values in the most sensitive region (50-300 Hz) of GW detectors. Thus, by extrapolating our results to lower frequency, we obtain a loss angle of $1.8 \times 10^{-4}$ at 100 Hz for the Ta$_2$O$_5$-ZrO$_2$ annealed at 700 $^\circ$C. This value has to be compared to $2.4 \times 10^{-4}$ for the Ta$_2$O$_5$-TiO$_2$ layers of GW detectors \cite{Granata20}.
\begin{table}
	\centering
	\begin{tabular}{ccc}
	    \hline
		$T_a$ [$^\circ$C] & $a$ [$10^{-4}$ rad Hz$^{-b}$] & $b$\\ \hline
		0 & 4.4 $\pm$ 0.4 & 0.05 $\pm$ 0.01\\
		500 \textcelsius & 1.6 $\pm$ 0.2 & 0.13 $\pm$ 0.01\\
		600 \textcelsius & 1.7 $\pm$ 0.2 & 0.12 $\pm$ 0.01\\
		700 \textcelsius & 0.8 $\pm$ 0.1 & 0.18 $\pm$ 0.02\\ \hline
	\end{tabular}
	\caption{Best-fit parameters of a power-law internal friction model $\varphi_c(f) = af^b$ for data sets of IBS Ta$_2$O$_5$-ZrO$_2$ thin films (LMA) measured on a disk suspended with a GeNS system, as function of annealing peak temperature $T_a$ (as-deposited coating samples are denoted by $T_a = 0$ $^\circ$C), shown in Figure \ref{FIG_LMA_coatLoss}. Each annealing treatment was performed in air and lasted 10 hours.}
	\label{TAB_LMAfit}
\end{table}

\subsection{ECR-IBD}
The coating mechanical loss at ambient temperature was characterized using fused-silica cantilevers. They were mounted in a stainless steel clamp within a vacuum chamber with residual pressure pressure below $5 \times 10^{-6}$~mbar, and excited using an electrostatic actuator~\cite{REID2006205}. The freely-decaying amplitude of the resonant motion was monitored using an optical shadow sensor. We then used Eqs. (\ref{EQ_coatLoss}) and (\ref{EQ_dilFact_cantilev}) to calculate the coating loss for the cantilever bending modes ~\cite{5391170}, and the results are presented in Section \ref{SECT_mech_compar}.

\begin{figure}[h]
\centering
	\includegraphics[width=0.85\textwidth]{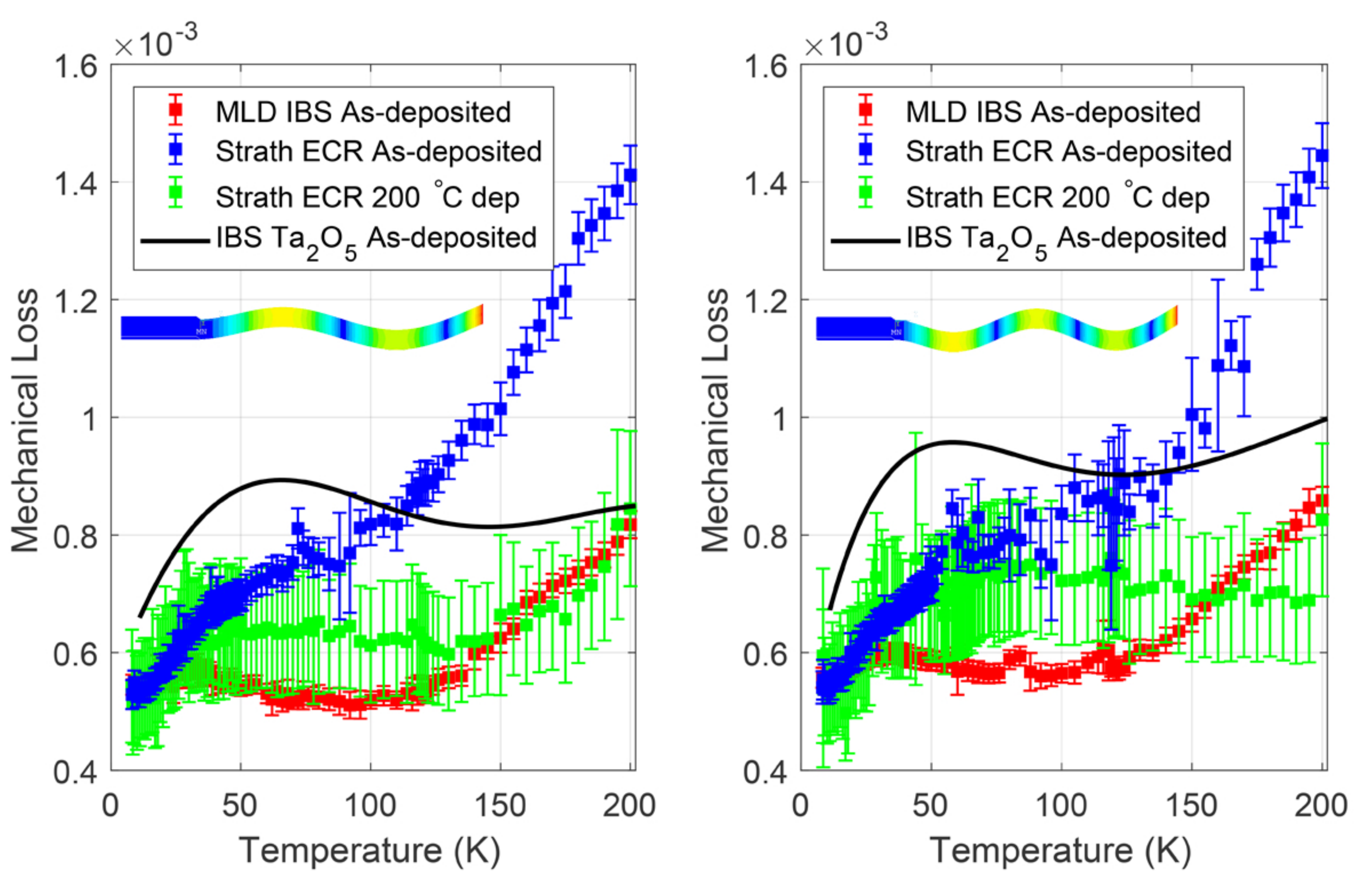}
	\caption{Cryogenic mechanical loss of as-deposited ECR-IBD Ta$_2$O$_5$-ZrO$_2$ coatings as a function of sample temperature, as measured on clamped $\sim$65-$\mu$m thick silicon cantilevers: (a) results for the third-order bending mode at $\sim$1.4 kHz, (b) results for the fourth-order bending mode at $\sim$2.7 kHz. Data of an ECR-IBD sample grown through standard deposition (blue markers) and that of an ECR-IBD grown on a substrate held at 200 $^{\circ}$C (green markers) are shown, and compared to the data of as-deposited IBS Ta$_2$O$_5$-ZrO$_2$ MLD2018 (see Figure \ref{FIG_MLD_cantilev_cryoT}) and Ta$_2$O$_5$ \cite{Martin_2010} coating samples. Finite-element representations of the measured resonant mode are visible in each plot.}
	\label{FIG_ECR_cantilev_cryoT}
\end{figure}

The same method was applied to perform cryogenic loss measurements of coated silicon cantilevers, whose results are shown in Figure \ref{FIG_ECR_cantilev_cryoT} for the the third- and fourth-order bending modes at $\sim$1.4 kHz and $\sim$2.7 kHz, respectively, and compared to the results obtained for the IBS sample MLD2018 (see Section \ref{SEC_MLDmech} and Figure \ref{FIG_MLD_cantilev_cryoT}). Below $\sim$100 K, both the ECR-IBD and IBS as-deposited Ta$_2$O$_5$-ZrO$_2$ coatings have a lower coating loss than that of as-deposited IBS Ta$_2$O$_5$ \cite{Martin_2010}. This shows that, prior to annealing, the addition of ZrO$_2$ already reduced the coating loss at low temperature. Below 30 K, the loss of the ECR-IBD coatings is comparable to that of IBS coatings.

Figure \ref{FIG_ECR_cantilev_cryoT} also shows the data of coating loss for an ECR-IBD Ta$_2$O$_5$-ZrO$_2$ coating sample grown on a substrate held at 200 $^{\circ}$C. By comparing the ECR-IBD samples, it can be seen that heating the substrate to a higher temperature had a considerable effect on the low-temperature coating loss, between 100 and 200 K: for instance, the coating loss of the sample grown at 200 $^{\circ}$C is 1.8 times smaller at 200 K. Finally, we found that the coating loss of the ECR-IBD sample grown at 200 $^{\circ}$C is similar to that of the IBS sample MLD2018 in its as-deposited state. 

Coating Young modulus $Y$ was measured using a Bruker Hysitron TriboScope nano-indenter system, with a three-sided Berkovich pyramid diamond tip ($Y_\textrm{\footnotesize{indenter}} = 1140$~GPa, and $\nu_\textrm{\footnotesize{indenter}} = 0.007$) and found to be $176 \pm 8$~GPa, assuming a coating Poisson ratio of $\nu = 0.297$ determined by molecular dynamics models \cite{Trinastic}. Measurements were taken on as-deposited thin films grown on 1-mm thick JGS-1 fused-silica witness samples, following the Oliver–Pharr method \cite{oliver_pharr_1992}.

\subsection{RBTD}
\label{SEC_RBTDmech}
Coating mechanical loss, Young modulus and Poisson ratio were estimated by measuring suspended coated fused-silica disks. The disks were measured by the LIGO Laboratory group at the California Institute of Technology, in a setup with four independent GeNS systems \cite{Cesarini09} installed in the same vacuum chamber to simultaneously measure all the accessible resonant modes of four samples \cite{Vajente17}.

Prior to coating deposition, the resonant frequencies and mechanical loss of the disks were characterized through specific measurements, which served as a reference and for background noise estimation for the subsequent series of measurements performed after deposition. All surfaces of the bare disks, including the edge, were polished to a standard optical grade to reduce the contribution of substrate surface losses \cite{penn_losses,Cagnoli18}.

After coating deposition, the disks were measured again, and then repeatedly annealed at increasing peak temperature $T_a$, from 300 to 800 $^{\circ}$C. After each annealing treatment, the resonant frequencies and mechanical loss of the disks were measured. Knowing the coating thickness and density from the previous characterizations (see Sections \ref{SECT_samples} and \ref{SECT_opt}), it was possible to compute the coating elastic properties, i.e. Young modulus $Y$ and Poisson ratio $\nu$, with a procedure similar to what explained in \cite{Li14} and in \cite{Vajente20} and to what was done in analyzing the LMA sample. The measured elastic properties, with their estimation uncertainties, were then used to compute the dilution factors $D$  and extract the coating loss angle according to Eq. (\ref{EQ_coatLoss}). The results are presented in Table \ref{TAB_mech_results} and Figures \ref{FIG_averaged-loss-angle-doping} and \ref{FIG_averaged-loss-angle-annealing}, and discussed in Section \ref{SECT_mech_compar}.

\subsection{HiPIMS/RF-MS}\label{SECT_MSmech}
Coating mechanical loss, Young modulus $Y$ and Poisson ratio $\nu$ were measured on suspended coated fused-silica disks by the LIGO Laboratory group at the California Institute of Technology, using the same setup and analysis methods described in Section \ref{SEC_RBTDmech}.

HiPIMS coating samples with different cation ratios $\eta=$ Zr/(Zr+Ta) were produced, in order to determine how ZrO$_2$ content affects the loss angle. Also, as mentioned in Section \ref{SECT_opt_compar}, RF-MS coating samples were grown with the same Ar:O$_2$ gas ratio, which had higher oxygen content and hence lower densities and refractive index. Samples were then annealed at increasing peak temperature $T_a$ by steps of 50 and 100 $^{\circ}$C, starting from 400 $^{\circ}$C, until they showed signs of crystallization. The annealing time for each soaking temperature $T_a$ was 10 hours. Depending on the cation ratio, the onset of crystallization was observed at temperatures which ranged from $T_a= 700$ $^{\circ}$C to $T_a = 850$ $^{\circ}$C.

\begin{figure}
\centering
    \includegraphics[width=9cm]{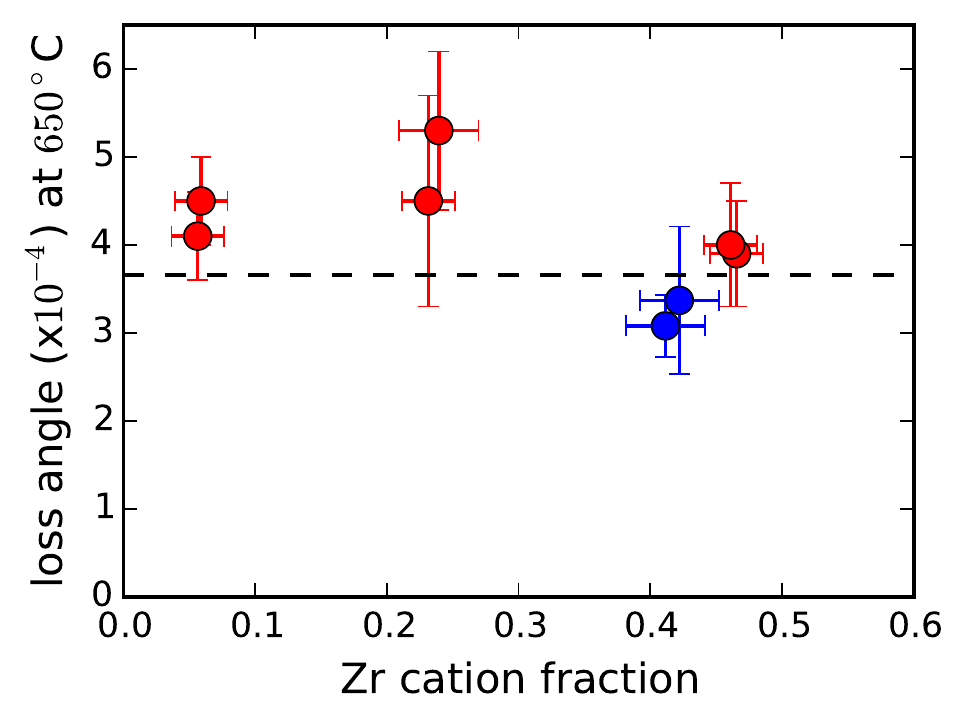}
    \caption{Coating loss of HiPIMS (red markers) and RF-MS (blue markers) Ta$_2$O$_5$-ZrO$_2$ thin films as a function of cation ratio $\eta =$ Zr/(Zr+Ta), as measured on suspended fused-silica disks after annealing at 650 $^{\circ}$C. The dashed black line represents the loss angle value of the annealed Ta$_2$O$_5$-TiO$_2$ layers of Advanced LIGO and Advanced Virgo \cite{Granata20}.}
    \label{FIG_MS_ZrVsLossAngle}
\end{figure}

Figure \ref{FIG_MS_ZrVsLossAngle} presents the coating loss angle as a function of the cation ratio $\eta$ at $650$ $^{\circ}$C, where it can be seen that most values are overlapping. This means that, within the uncertainties of the measurements, $\eta$ does not considerably affect the coating loss angle. Although still within measurement uncertainty, however, RF-MS samples feature a slightly lower coating loss angle. Therefore, even if HiPIMS samples were shown to be denser and closer to stoichiometry than RF-MS samples, this did not translate into a decrease of the coating loss angle.

\subsection{Comparative summary}
\label{SECT_mech_compar}
We discuss here the effects of different cation ratios $\eta =$ Zr/(Zr+Ta) and annealing peak temperatures $T_a$ on the mechanical loss of our coating samples, limiting our analysis to the results obtained at ambient temperature so that we can compare samples from all the growth techniques presented in this work.
%
%
To ease comparison, an average coating loss $\overline{\varphi}$ can be taken from every set of measured resonant modes of each sample.

\begin{figure}
	\centering
	\includegraphics[width=\textwidth]{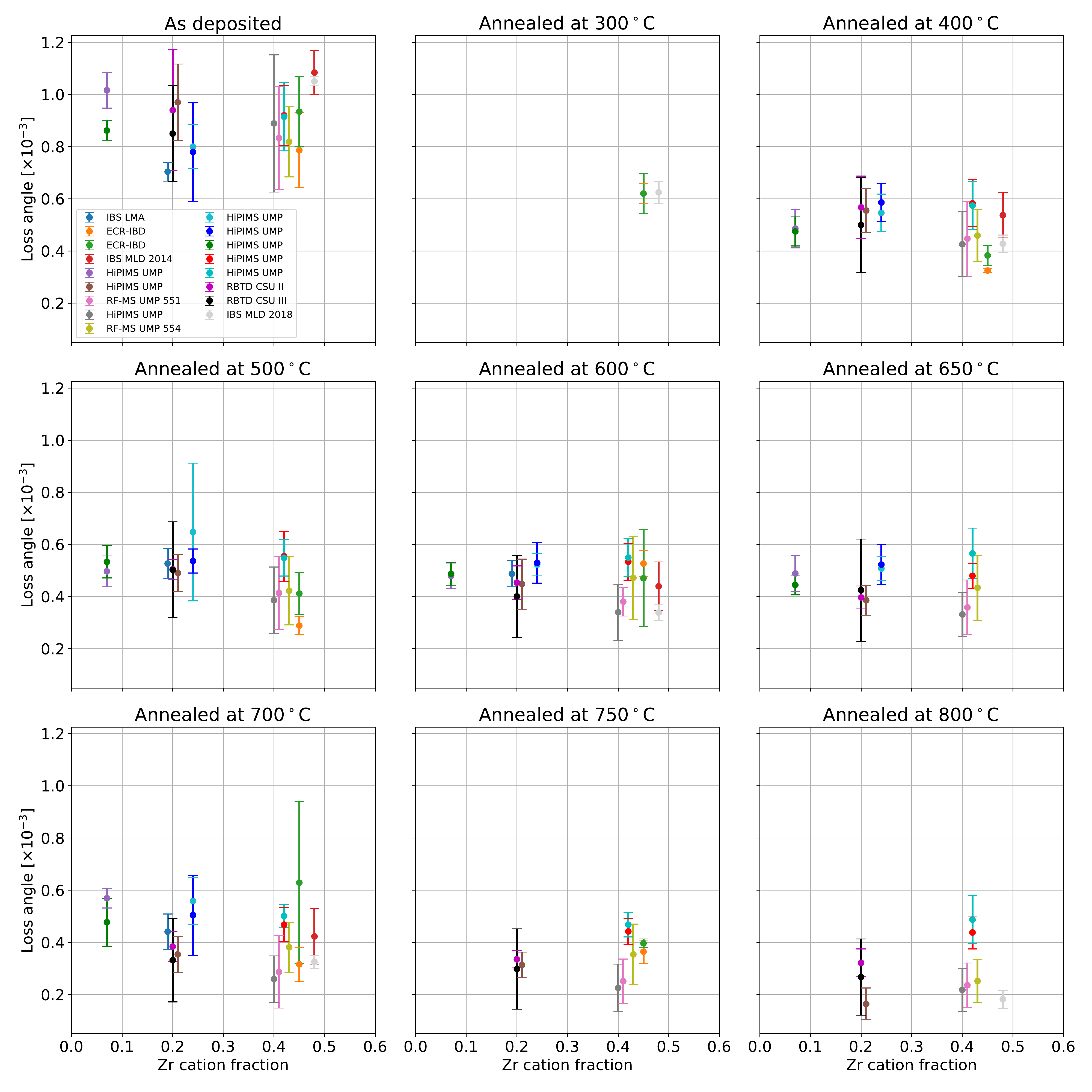}
	\caption{Frequency-averaged coating loss angle $\overline{\varphi}$ of all the Ta$_2$O$_5$-ZrO$_2$ coating samples considered in this work, as a function of cation ratio $\eta =$ Zr/(Ta+Zr). Each panel corresponds to a different annealing temperature $T_a$.}
	\label{FIG_averaged-loss-angle-doping}
\end{figure}

Figure \ref{FIG_averaged-loss-angle-doping} shows the resulting $\overline{\varphi}$ as a function of cation ratio $\eta =$, where it can be seen that $\eta =$ does not affect significantly the coating mechanical loss. Furthermore, coating loss seems to be fairly independent of the technique used to grow the samples.

The samples were progressively annealed through steps of increasingly high peak temperature $T_a$ to test the effect of annealing on coating loss. For these experiments, the fused silica substrate is typically annealed to temperatures higher than 900 $^\circ$C before being coated. The high temperature annealing of the bare substrate lowers its mechanical loss ($\varphi_s < 10^{-7}$) such that it is negligible compared to the coating loss. In addition, the substrate loss will not change with lower temperature annealing cycles, so that any change in the coated sample loss may be attributed solely to the coating.

\begin{figure}
	\centering
	\includegraphics[width=\textwidth]{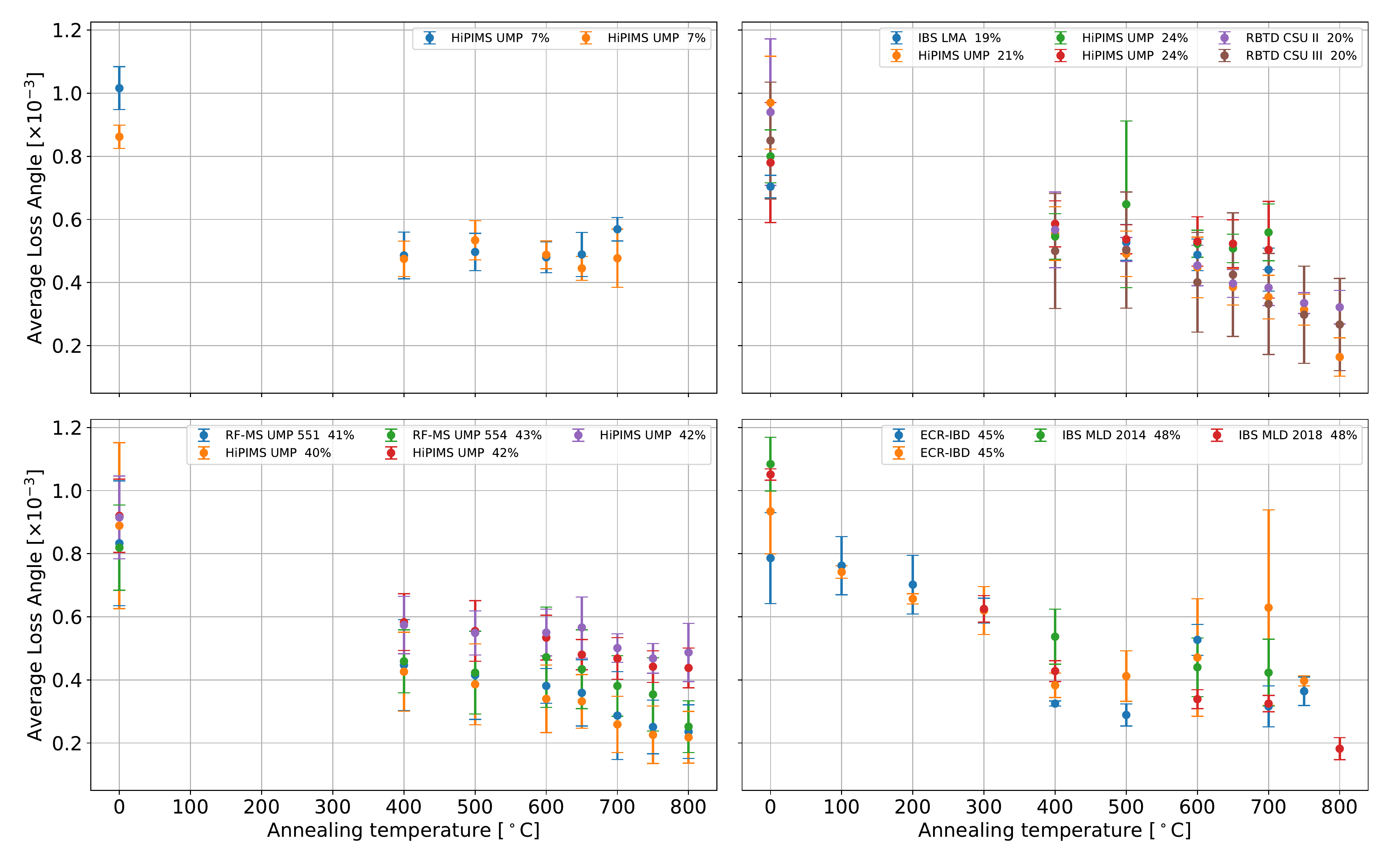}
	\caption{Frequency-averaged coating loss angle $\overline{\varphi}$ of all the Ta$_2$O$_5$-ZrO$_2$ coating samples considered in this work, as a function of the annealing temperature $T_a$ (as-deposited samples are denoted by $T_a = 0$ $^{\circ}$C). The four panels group together different samples with similar cation ratio $\eta =$ Zr/(Ta+Zr).}
	\label{FIG_averaged-loss-angle-annealing}
\end{figure}

Figure \ref{FIG_averaged-loss-angle-annealing} shows that annealing significantly reduced the coating mechanical loss, from about 10$^{-3}$ for as-deposited samples down to $(2-4) \times 10^{-4}$ for samples treated at 800 $^{\circ}$C.
%
\section{Effects of annealing}
\label{SECT_annealing}
\begin{figure}
\centering
\includegraphics[width=9cm]{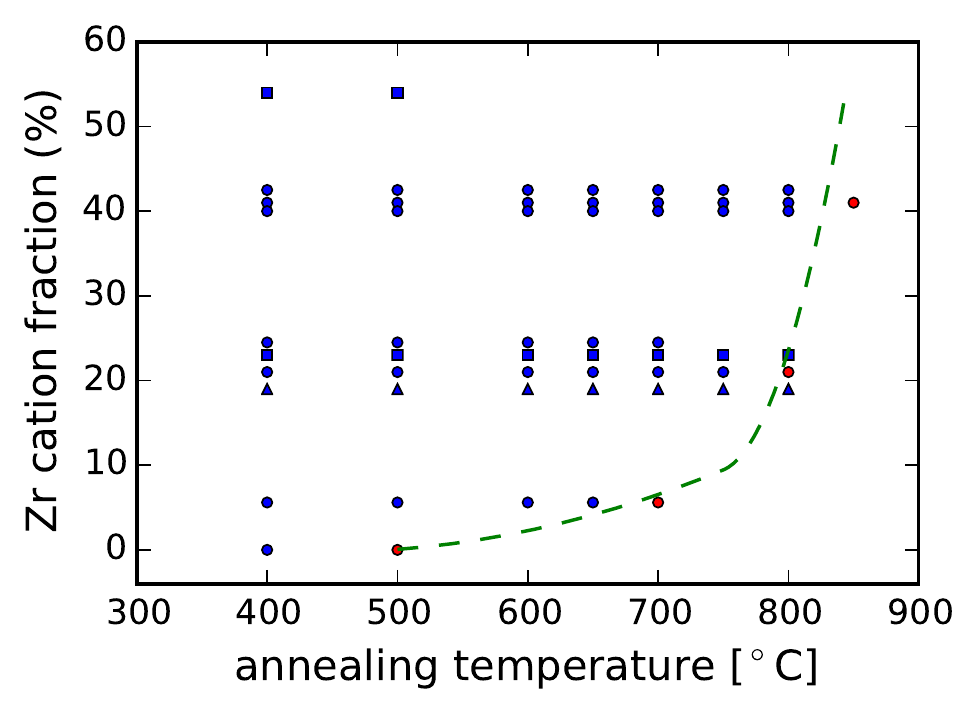}
\caption{State of IBS, RBTD and HiPIMS/RF-MS Ta$_2$O$_5$-ZrO$_2$ thin films after annealing at temperature $T_a$, as a function of cation ratio $\eta =$ Zr/(Ta+Zr). The blue markers denote amorphous samples, while red ones indicate that the transition to a poly-crystalline phase was observed. The dashed curve shows the approximate limit of the phase transition as a function of $\eta$. 
}
\label{FIG_crystMap}
\end{figure}

As expected, co-sputtering zirconia proved to be an efficient way to frustrate crystallization in tantala thin films, allowing for a substantial increase of the maximum annealing temperature $T_a$. Figure \ref{FIG_crystMap} shows that higher cation ratios $\eta =$ Zr/(Zr+Ta) allowed to increase $T_a$ up to 800 $^{\circ}$C, without crystallization occurring. More specifically, IBS samples have been tested up to 800 $^{\circ}$C and no sign of crystallization was observed. RBTD sample CSU II ($\eta = 0.23$) was amorphous after treatment at 800 $^{\circ}$C. HiPIMS/RF-MS samples showed signs of crystallization for temperatures ranging from 700 to 850 $^{\circ}$C, depending on the cation ratio $\eta$.

Figures \ref{FIG_averaged-loss-angle-doping} and \ref{FIG_averaged-loss-angle-annealing} show that annealing significantly reduced the coating mechanical loss, from about $\overline{\varphi}\sim10^{-3}$ for the as-deposited samples down to $\overline{\varphi}\sim(2-4)\times10^{-4}$ for the samples treated at 800 $^{\circ}$C, whereas, within the measurement uncertainties, the doping does not affect the loss angle for a given $T_a$.

However, cracks appeared on IBS (MLD), RBTD and HiPIMS/RF-MS samples after annealing at 500 $^{\circ}$C, and on IBS (LMA) samples after annealing at 700 $^{\circ}$C. Likely, this happened because of a mismatch between the thermal expansion coefficients of the thin films and substrates, causing high thermal stress. Although in principle our mechanical loss measurements are sensitive to such defects, we found no evidence that our results were affected. At least for some IBS samples (LMA), we found that the formation of cracks could be avoided by decreasing the annealing heating and cooling rates down to 7 $^{\circ}$C/h.

Nevertheless, starting from $T_a = 500$ $^{\circ}$C, bubble-like defects of different number and size, detected with an optical microscope, appeared at variable depth in the IBS HR coating samples, likely due to the presence of incorporated argon within the layers \cite{cevro95,fazio2020structure,paolone2021}. We observed no trace of such defects in the annealed IBS single-layer annealed samples deposited under strictly identical conditions, indicating that this phenomenon only occurs when layers are stacked. It is not clear how to avoid this issue yet, and we are currently working to solve this problem.

At the same time, annealing at $T_a = 500$ $^{\circ}$C decreased the optical absorption of the IBS HR samples down to 0.5 ppm.
%
\section{Raman spectroscopy}
\label{SECT_Raman}
Raman spectroscopy probes the vibrational modes of a material constituents. In previous studies on amorphous silicon (a-Si), Roorda et al. found a correlation between structural relaxation and the width of a peak in its Raman spectrum \cite{RoordaStrucural}. A sharpening of the peak was interpreted as the results of a decrease in the bond angle distribution upon annealing. Granata et al. also found a correlation between the mechanical loss and the normalized area of a Raman line in IBS and bulk silica \cite{Granata18}. Thus, it is of interest to see if there is a correlation between the Raman spectra of Ta$_2$O$_5$-ZrO$_2$ thin films and their loss angle, via structural relaxation. 

\begin{figure}
\centering
\includegraphics[width=8cm]{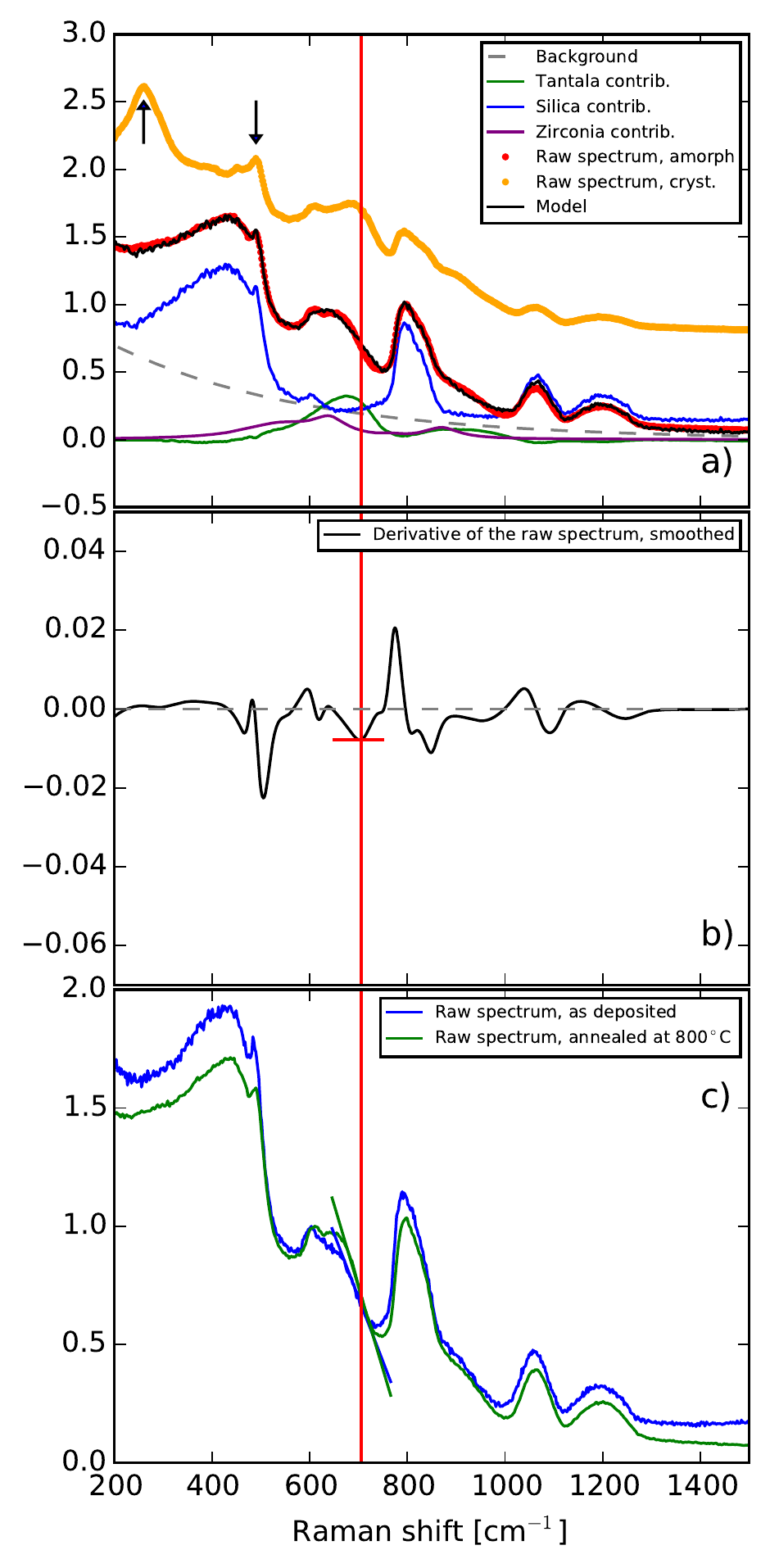}\hfill
\caption{Measured Raman spectra of HiPIMS/RF-MS Ta$_2$O$_5$-ZrO$_2$ thin films and their analysis: a) example of a spectrum obtained for an amorphous sample (red line), with superimposed a model (black) resulting from different contributions summed together (blue, green, dashed grey, see text for details); b) derivative of the raw Raman spectrum, smoothed using bi-cubic splines; c) raw, normalized spectrum of the same sample, as-deposited (blue) and annealed (green). The slopes extracted using the derivative method shown in b) are represented by straight lines. The vertical red line across all panels is the position of the minimum of the first derivative of the Zr contribution (purple curve) shown in a).}\label{FIG_ramanDerivative}
\end{figure}

Raman spectroscopy was performed on the coated silica disks with a Renishaw Invia Reflex microscope, with an Ar laser at a wavelength of 514 nm. The measurement was performed three times at different locations on each sample, then the spectra of each sample were averaged. It was carried out on all the MS as-deposited samples and on seven MS samples after annealing at 800 $^{\circ}$C. Note that two of these annealed samples showed signs of crystallization, with significant change in the coating Young modulus and loss angle measurements (see Table \ref{TAB_mech_results} and Sections \ref{SECT_MSmech} and \ref{SECT_mech_compar}). Figure~\ref{FIG_ramanDerivative}a presents an example of the Raman spectrum measured on a crystallized sample, where the arrows identify peaks associated with the crystal phase \cite{PEREZ2017303}.

Tantala-zirconia has a complex structure, which translates into several peaks in the Raman spectrum, some of them overlapping, and makes the data analysis non-trivial. In addition, Ta$_2$O$_5$-ZrO$_2$ thin films are transparent at 514 nm, so that peaks from the underlying silica substrate also contribute to the measured spectra. Therefore, the initial step of the analysis was to decorrelate the contribution of the thin film from that of the substrate. Figure~\ref{FIG_ramanDerivative}a shows a Raman spectrum of an amorphous sample (red curve), with superimposed a model (black) resulting from different contributions summed together: (i) the spectrum of a bare silica substrate (blue), (ii) the spectrum of a pure tantala thin film (green) obtained from a previous study \cite{Vajente_2018}, (iii) a background contribution represented by an exponential (dashed grey), (iv) additional contributions required so that the model correctly fit the data (purple), which are then assumed to be due to the addition of zirconia and which were modeled using four Lorentzian functions. Thanks to this model, we can identify where the thin film is the main contributor to the measured spectrum: referring to the purple curve of Figure~\ref{FIG_ramanDerivative}a, this can be seen to happen in the interval from 600 to 750 cm$^{-1}$. An additional contribution from the thin film is observed near 900 cm$^{-1}$, but is less clearly separated from the substrate signal and was disregarded in the following analysis.  

Next, we wanted to estimate the width of the Raman peak attributed to the thin film. However, trying to subtract substrate contributions or using a fit model would likely influence the shape of the peak, hence its width. To avoid this, we rather relied directly on the experimental data: we calculated the maximum slope of the raw signal on one side of the peak, at a location where the coating contribution to the spectrum is dominant. Indeed, the slope of a peak edge is a proxy for the peak width: for both Lorentzian and Gaussian functions, the maximum slope is proportional to the inverse squared width. Thus narrower peaks, which might be attributed to a more relaxed material, should feature a steeper slope on each side. To this end, the raw spectra were smoothed with a bi-cubic spline and normalized so that their maximum, in the 550 cm$^{-1}$ to 750 cm$^{-1}$ spectral region, where the contribution of the thin film is at the highest, is at unity. We then computed the derivative, an example of the result is found in Figure~\ref{FIG_ramanDerivative}b.

In all panels of Figure~\ref{FIG_ramanDerivative}, the position of interest, i.e. the point where the slope of the signal related to the Zr contribution is the steepest, is represented by a vertical red line. In Figure~\ref{FIG_ramanDerivative}c, we can see two raw spectra of the same sample, acquired before and after annealing. The slope computed using the spline fit at the Raman shift of interest is also plotted as a straight line, for each spectrum. We can observe from the raw data that the slope around 705 cm$^{-1}$ changes upon annealing and are well represented by the first derivative deduced from the method illustrated in Figure~\ref{FIG_ramanDerivative}b.
\begin{figure}
\centering
\includegraphics[width=12.0cm]{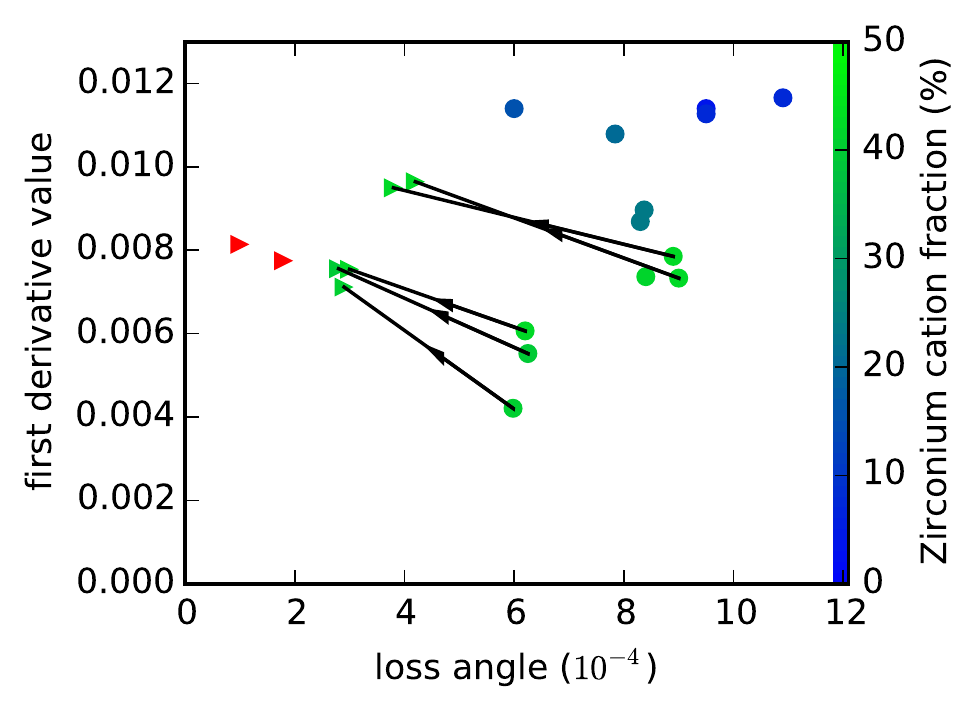}
\caption{Absolute value of the minimum of the derivative of the Raman spectrum near 705 cm$^{-1}$ as a function of the loss angle, for HiPIMS/RF-MS Ta$_2$O$_5$-ZrO$_2$ thin films. The color scale corresponds to the cation ratio $\eta =$ Zr/(Zr+Ta), circles denote spectra acquired on the as-deposited samples and triangles denote those acquired on the annealed samples, the same sample being linked by a black arrow. Red markers represent samples showing signs of crystallization.}\label{FIG_ramanDerivativeVsLossAngle}
\end{figure}
If Raman peaks sharpen with relaxation, and relaxation decreases the loss angle, we expect a higher value of the derivative with decreasing loss angle.

Figure~\ref{FIG_ramanDerivativeVsLossAngle} shows the absolute value of the first derivative around 705 cm$^{-1}$ as a function of the loss angle, where the color scale represent the cation ratio $\eta =$ Zr/(Zr+Ta). The red markers correspond to the two crystallized samples which show characteristic peaks in their Raman spectrum, and most probably some delamination, making their loss angle measurements unreliable. The circles are for measurements carried out on as-deposited samples while the triangles are for the annealed ones, and the same samples are linked by an arrow pointing toward the annealed one. 

Figure~\ref{FIG_ramanDerivativeVsLossAngle} clearly shows that, upon annealing, each individual sample sees its peak slope increase and its loss angle decrease at the same time. Hence, the Raman peak sharpens upon relaxation, similar to what is observed in a-Si \cite{RoordaStrucural}. 
%
However, if one considers only the subset of the as-deposited samples (circles), it appears that the lower the loss angle, the lower the slope, which means the Raman peaks are wider. Also, samples having the lowest loss angle are the ones with the highest cation ratio $\eta$ (greener data points). Hence, the shape of the Raman peak near 705 cm$^{-1}$  cannot be attributed solely to relaxation, as it is also clearly influenced by the micro-structural modifications that large amounts of Zr induce into the thin films, which in turn have an impact on the vibrational states distribution and hence on the shape of the Raman spectrum.

Therefore, these results show that Raman spectroscopy could be used as a probe to monitor the relaxation state of tantala-based alloys, and that relaxation is linked to the loss angle. Similar conclusions were reached using X-ray diffraction on tantala \cite{Vajente_2018}, where the X-ray peak of tantala sharpened with decreasing loss angle. However, the Raman peak shape is also influenced by the amount of doping, so that while it can be used to monitor the evolution of a specific sample, it would be difficult to use it more generally as a probe of the state of relaxation in a material or a predictor of the loss angle.
%
\section{Summary and Conclusions}
\label{SECT_discussion}
In order to decrease coating Brownian noise in current GW detectors, we developed a set of co-sputtered tantala-zirconia thin films. We extensively characterized these coatings for optical and mechanical properties, to test their performances against the very stringent requirements of GW detectors.

As expected, co-sputtering zirconia proved to be an efficient way to frustrate crystallization in tantala thin films, allowing for a substantial increase of the maximum annealing temperature $T_a$,  hence for a decrease of coating mechanical loss. The lowest average loss, $\overline{\varphi_c} = 1.8 \times 10^{-4}$, was measured on IBS sample MLD2018 with cation ratio Zr/(Ta+Zr) = 48.5 $\pm$ 0.4 \% annealed at 800 $^\circ$C. This value has to be compared to $\varphi_c = (2.3-3.4) \times 10^{-4}$ of tantala-titania layers in current GW detectors \cite{Granata20,gras18}. 

As measured on IBS HR samples with the same design of those of Advanced LIGO and Advanced Virgo, we observed as little optical absorption (0.5 ppm) as in current GW detectors \cite{Pinard:17,Degallaix19}. The same samples, however, featured an unusually high value of scattering, 45 ppm, compared to that of present detectors (5 ppm) \cite{Pinard:17,Degallaix19}.

For some samples, we measured a refractive index at 1064 and 1550 nm higher than that of pure tantala thin films. This implies that HR stacks containing zirconia-doped tantala layers could be even thinner than HR coatings of present detectors, further decreasing coating Brownian noise \cite{coatingR&Dreview}.

For use in future GW detectors, further development will be needed to decrease scattering and to avoid the formation of bubble-like defects upon annealing. We are currently working to find appropriate solutions to these issues.

\section*{Acknowledgments}
The research performed at the Laboratoire des Mat\'{e}riaux Avanc\'{e}s was partially supported by the Virgo Coating Research and Development (VCR\&D) Collaboration.

The work performed at U. Montr\'{e}al and \'{E}cole Polytechnique de Montr\'{e}al was supported by the Natural Sciences and Engineering Research Council of Canada (NSERC), the Canadian foundation for innovation (CFI) and the Fonds de recherche Qu\'{e}bec, Nature et technologies (FQRNT) through the Regroupement Qu\'{e}b\'{e}cois sur les mat\'{e}riaux de pointe (RQMP). These authors thank S. Roorda, M. Chicoine, L. Godbout and F. Turcot for fruitful discussions and technical support.

The work performed at the University of Glasgow was supported by the Science and Technology Facilities Council (STFC, ST/N005422/1 and ST/I001085/1) and the Royal Society (UF100602 and UF150694).

The research performed at Hobart and William Smith Colleges was supported by National Science Foundation grant awards PHY-1307423, 1611821, and 1707863.

\section*{References}
\bibliography{zirconia}
\bibliographystyle{plain}

\end{document}